\documentclass[twocolumn]{article}

\usepackage{amsmath}
\usepackage{graphicx}
\usepackage[utf8]{inputenc}
\usepackage[T1]{fontenc}
\usepackage{bm}
\usepackage{braket}
\usepackage{appendix}
\usepackage{authblk}
\usepackage{booktabs}

\author[1]{R.~Laasner}
\author[1]{W.~Huhn}
\author[2]{J.~Colell}
\author[2]{T.~Theis}
\author[1]{V.~W.~Yu}
\author[2,3]{W.~Warren}
\author[1,2]{V.~Blum}
\affil[1]{Department of Mechanical Engineering and Materials Science, Duke University, Durham NC 27708, United States}
\affil[2]{Department of Chemistry, Duke University, Durham NC 27708, United States}
\affil[3]{Department of Physics, Duke University, Durham NC 27708, United States}

\title{Molecular NMR shieldings, \textit{J}-couplings, and magnetizabilities from numeric atom-centered orbital based density-functional calculations}

\date{}

\begin{document}

\maketitle

\begin{abstract}
  We describe an accurate and scalable implementation for the computation of molecular nuclear magnetic resonance shieldings, \textit{J}-couplings, and magnetizabilities within nonrelativistic semilocal density functional theory, based on numeric atom-centered orbital (NAO) basis sets. We compare the convergence to the basis set limit for two established types of NAO basis sets, called NAO-VCC-$n$Z and FHI-aims-09, to several established Gaussian-type basis sets. The basis set limit is reached faster for the NAO basis sets than for standard correlation consistent Gaussian-type basis sets (cc-pV$n$Z, aug-cc-pV$n$Z, cc-pCV$n$Z, aug-cc-pCV$n$Z). For shieldings, the convergence properties and accuracy of the NAO-VCC-$n$Z basis sets are similar to Jensen's polarization consistent (pc) basis sets optimized for shieldings (pcS-$n$). For \textit{J}-couplings, we develop a new type of NAO basis set (NAO-J-$n$) by augmenting the NAO-VCC-$n$Z basis sets with tight s-functions from Jensen's pcJ-$n$ basis sets, which are optimized for \textit{J}-couplings. We find the convergence of the NAO-J-$n$ to be similar to the pcJ-$n$ basis sets. Large scale applicability of the implementation is demonstrated for shieldings and \textit{J}-couplings in a system of over 1,000 atoms.
\end{abstract}

\section{Introduction}

Nuclear magnetic resonance (NMR) spectroscopy is one of the most widely used experimental techniques for probing the structure and dynamics of materials, reactions and morphology. It has found widespread application in chemistry, medicine and geophysics \cite{markwick08},\cite{marion13},\cite{levitt08},\cite{freedberg14},\cite{lesage09},\cite{ashbrook09},\cite{brown01},\cite{sarracanie15},\cite{song12}. The NMR chemical shifts and the indirect dipole-dipole couplings (\textit{J}-couplings) are highly sensitive to the electronic structure and can serve as accurate probes of the local chemical environment \cite{darbeau06},\cite{kwan11}. Due to the often empirical nature of interpreting experimental NMR data, first-principles calculations, particularly density functional theory (DFT) ones, are widely used to help assign the chemical shifts and \textit{J}-couplings to specific nuclei \cite{mulder10},\cite{thonhauser09},\cite{helgaker00},\cite{cheeseman96},\cite{autschbach00a},\cite{autschbach00b},\cite{autschbach04},\cite{helgaker08},\cite{joyce07},\cite{autschbach09}. Additionally, first-principles derived NMR parameters may serve as fundamental input to model Hamiltonian descriptions \cite{hogben11},\cite{edwards14},\cite{guduff17} of the statistical mechanics and time evolution of nuclear spin systems, as probed, for instance, in magnetic resonance imaging techniques.

One of the most important numerical factors influencing the accuracy of calculated NMR parameters is the quality of the quantum-mechanical basis set, especially in the region near the nucleus. All-electron implementations, which treat the potential in this region without \textit{a priori} approximations, exist for several widely used basis set types such as Gaussian type orbitals (GTOs) \cite{helgaker91},\cite{ruud94},\cite{cheeseman96},\cite{sychrovsky00},\cite{ochsenfeld11}, linearized augmented plane waves (LAPW) \cite{laskowski12}, Slater-type orbitals (STOs) \cite{schreckenbach95},\cite{skachkov10}, and wavelets \cite{jensen16}. Close to all-electron accuracy has also been reported with projector augmented waves (PAW) \cite{pickard01},\cite{joyce07},\cite{wijs17} and ultrasoft pseudopotentials \cite{yates07}.

The purpose of the present work is to investigate the use of numeric atom-centered orbitals (NAOs) \cite{blum09},\cite{zhang13},\cite{delley90},\cite{koepernik99},\cite{kenny09} as all-electron basis sets for three magnetic response properties: NMR shielding tensors (and chemical shifts), magnetizabilities, and \textit{J}-couplings. NAO basis functions can be expressed in the form
\begin{equation}
  \label{eq:nao_def}
  \phi_i(\bm r) = Y_{lm}(\Omega) \frac{u_i(|\bm r - \bm R_i|)}{|\bm r - \bm R_i|}
\end{equation}
in terms of the spherical harmonics $Y_{lm}$, where $l$ and $m$ are the angular momentum quantum numbers, and the radial functions $u_i(r)$, where $i$ indexes different atoms, radial functions, and angular momenta $l$. $\bm r$ is the position vector, $\bm R_i$ is the origin of the basis function $i$, and $\Omega$ is the angular part of the position vector referenced to $\bm R_i$. The radial functions $u_i(r)$ are numerically tabulated and evaluated as cubic splines; therefore, their shape is fully flexible. Importantly, the correct shape close to the nucleus can be captured with high accuracy for the density functional employed in a specific simulation.

We compare the performance of NAO basis sets to several Gaussian type basis sets that are well established in the community. In particular, we test three types of NAO basis sets, referred to as ``FHI-aims-09'' \cite{blum09}, NAO-VCC-$n$Z \cite{zhang13}, and NAO-J-$n$ (introduced in the present work), against a group of GTO basis sets: cc-pV$n$Z \cite{dunning89}, aug-cc-pV$n$Z \cite{kendall92}, cc-pCV$n$Z and aug-cc-pCV$n$Z \cite{woon95}, (aug-)pc-$n$ \cite{jensen01}, (aug-)pcS-$n$ \cite{jensen08}, and (aug-)pcJ-$n$ \cite{jensen06}. The main attributes of each basis set are summarized in Table~\ref{tab:basis_sets} and are described in more detail in the following.
\begin{table*}
  \centering
  \caption{Basis sets tested in this work. ``Type'' refers to either numeric atom centered orbitals (NAOs) or Gaussian type orbitals (GTOs). To characterize the size of different basis sets, $N_b$ is the number of basis functions averaged over the basis size of isolated B, Si, C, P, N, S, O, and Al atoms (not all basis sets are available for the other elements).$N_b = (N^{B}+N^{Si}+N^{C}+N^{P}+N^{N}+N^{S}+N^{O}+N^{Al})/8$, where $N^X$ is the basis size of element $X$. $N_b^{\text{m}}$ and $N_b^{\text{M}}$ are the minimum and maximum $N_b$ values for the given basis set type.}
  \begin{tabular}{lllll}
    \toprule
    Basis set & Type & Levels & $N_b^{\text{m}}$--$N_b^{\text{M}}$ & Target \\
    \midrule
    FHI-aims-09 \cite{blum09} & NAO & 1,2,3,4 & 21--86 & General purpose DFT \\
    NAO-VCC-$n$Z \cite{zhang13} & NAO & 2,3,4,5 & 23--100 & General purpose with focus on sums over unoccupied \\
    &&&& states for perturbation theory (RPA or MP2) \\
    NAO-J-$n$ & NAO & 2,3,4,5 & 27--105 & \textit{J}-couplings \\
    cc-pV$n$Z \cite{dunning89} & GTO & D,T,Q,5,6 & 16--142 & General purpose \\
    aug-cc-pV$n$Z \cite{kendall92} & GTO & D,T,Q,5,6 & 25--191 & General purpose and specifically electron affinities \\
    cc-pCV$n$Z \cite{woon95} & GTO & D,T,Q,5,6 & 23--253 & General purpose with focus on core states\\
    aug-cc-pCV$n$Z \cite{woon95} & GTO & D,T,Q,5 & 32--199 & General purpose with focus on core states \\
    pc-$n$ \cite{jensen01} & GTO & 0,1,2,3,4 & 11--107 & General purpose \\
    aug-pc-$n$ \cite{jensen01} & GTO & 0,1,2,3,4 & 15--143 & General purpose \\
    pcS-$n$ \cite{jensen08} & GTO & 0,1,2,3,4 & 11--122 & Shieldings \\
    aug-pcS-$n$ \cite{jensen08} & GTO & 0,1,2,3,4 & 15--158 & Shieldings \\
    pcJ-$n$ \cite{jensen06} & GTO & 0,1,2,3,4 & 15--130 & \textit{J}-couplings \\
    aug-pcJ-$n$ \cite{jensen06} & GTO & 0,1,2,3,4 & 19--177 & \textit{J}-couplings \\
    \bottomrule
  \end{tabular}
  \label{tab:basis_sets}
\end{table*}

The FHI-aims-09 basis sets are general-purpose basis sets that were created for DFT-based total-energy calculations, covering 102 chemical elements. For each chemical element, they are grouped into tiers, or levels, of basis functions, leading to a series of basis sets of increasing accuracy. In order to construct this basis set library, successive basis functions were selected by minimizing the total energies of non-spinpolarized symmetric dimers using hydrogen-like, cation-like, or atom-like functions with a variable confinement potential (see Ref.~\cite{blum09} for details). Those functions that led to the largest decrease in the total energies were added to the basis set. By convention, each FHI-aims-09 basis set includes the occupied core and valence functions of spherical free atoms as the ``minimal basis.'' A ``tier 1'' basis set retains this minimal basis and adds a set of additional basis functions. A ``tier 2'' basis set adds a second ``tier'' of basis functions to the tier 1 basis set, and so on. The FHI-aims-09 basis sets have proven to be accurate basis sets for ground state all-electron calculations \cite{lejaeghere16}. Accuracy on the order of a few meV/atom or better for the total energy was demonstrated in reference to multiresolution wavelets \cite{jensen17}. However, only the DFT total energy was considered in the construction of the FHI-aims-09 basis sets. The response of the density to external perturbations was not specifically considered. Thus, it is not \textit{a priori} clear how well they will capture second order properties such as the magnetic response parameters.

The valence-correlation consistent NAO-VCC-$n$Z \cite{zhang13} basis sets for elements H-Ar (1-18) were constructed with a similar strategy as Dunning's \cite{dunning89} GTO cc-pV$n$Z basis sets. They were designed to capture the density response of the valence electrons of small molecules systematically and were shown to be accurate for total energy calculations that contain explicit sums over unoccupied states (e.g., second-order Møller-Plesset theory \cite{moller34} or the random-phase approximation \cite{ren12},\cite{ren12a}). Thus, they are expected to be accurate for quantities such as the paramagnetic contribution to a magnetic property, which also involves sums over unoccupied states.

In this work, we introduce a new group of hybrid basis sets, ``NAO-J-$n$'', which are designed specifically for \textit{J}-couplings. They combine the NAO-VCC-$n$Z with highly localized Gaussian s-type functions taken from Jensen's pcJ-$n$ basis sets \cite{jensen06} (see below), which were optimized for \textit{J}-couplings.

For comparison and reference values, we consider the following GTO basis sets. The cc-pV$n$Z basis sets are general-purpose basis sets, first proposed by Dunning \cite{dunning89}, and designed by successively constructing levels of basis functions by minimizing the total energies of isolated atoms relative to multiconfiguration self-consistent-field calculations. The aug-cc-pV$n$Z \cite{kendall92} basis sets add long ranged functions, so-called augmentation functions, to the cc-pV$n$Z basis sets. They were originally developed by targeting electron affinities specifically, but are more generally expected to increase the accuracy of calculations of weakly bound and excited states. The cc-pCV$n$Z and aug-cc-pCV$n$Z basis sets add additional Gaussian functions to better capture the correlation effects of the core electrons \cite{woon95}.

We also investigate another class of GTOs --- the general-purpose polarization consistent (pc-$n$ and aug-pc-$n$) basis sets developed by Jensen \cite{jensen01} and their respective optimized versions for shieldings and $J$-couplings. The pc-$n$ basis sets were built by successively adding polarization functions to a minimal set, but used a different construction strategy than the cc-pV$n$Z basis sets. Instead of focusing on minimizing the energy of isolated atoms, the pc-$n$ basis sets target the total Hartree-Fock energy of molecular systems. Building on top of the pc-$n$ basis sets, the pcS-$n$ \cite{jensen08} and pcJ-$n$ \cite{jensen06} basis sets were developed, which are optimized for NMR shieldings and \textit{J}-couplings, respectively. The pcS-$n$ and pcJ-$n$ basis sets perform very well for their respective purposes and members of these basis sets serve as the point of reference in our work.

In this paper, we present an implementation of the nonrelativistic formalism of NMR shieldings, magnetizability, and \textit{J}-couplings, applicable to molecules in semilocal DFT, in the NAO based all-electron code FHI-aims \cite{blum09},\cite{ren12},\cite{havu09},\cite{ihrig15}. For magnetizabilities, the NAOs show faster convergence than any of the tested GTO basis sets. For shieldings, the NAOs are on par with the pcS basis sets. For \textit{J}-couplings, the convergence of the new NAO-J-$n$ basis sets is similar to the pcJ-$n$ basis sets. For larger molecules, we also comment on the numerical stability of all investigated quantities related to ill-conditioning \cite{moncrieff05} as the basis set size increases towards the basis set limit. For the NAO basis sets and systems investigated here, we do not observe stability issues, but ill-conditioning appears to limit the accuracy reachable with the larger ``augmented'' GTO basis sets for large systems.

Finally, we demonstrate the applicability of our implementation for large scale systems by calculating shieldings and \textit{J}-couplings in a test system of 1,052 atoms. The test system is a DNA segment that was used by the authors of Ref.~\cite{ochsenfeld11} in a past validation of an NMR implementation.

\section{Implementation}

\subsection{Definitions}

The NMR chemical shielding tensor is defined as the proportionality between an external magnetic field $\bm B$ and the induced magnetic field $\bm B_{\text{ind}}$ at the position of a nucleus:
\begin{equation}
  \overleftrightarrow\sigma_A = -\bm B_{\text{ind}}(\bm R_A)/\bm B ,
\end{equation}
  where $\leftrightarrow$ is the tensor notation. The ratio of induced and external fields is not necessarily linear in the definition of $\sigma_A$, but all experimentally available magnetic fields are sufficiently small that the relation can be considered linear to very good accuracy. The chemical shielding tensor can be identified with the second derivative of the total energy with respect to the applied magnetic field $\bm B$ and a nuclear magnetic moment  $\bm\mu_A$ \cite{helgaker99}:
\begin{equation}
  \label{eq:shield_def}
  \overleftrightarrow\sigma_A = \left. \frac{\partial ^2 E}{\partial \bm B\partial \bm\mu_A} \right|_{\bm B=0;\bm\mu_A=0},
\end{equation}
where $\bm\mu_A$ is placed at the position of the NMR active nucleus. The energy expression in Eq.~\eqref{eq:shield_def} does not contain the classical Zeeman term \cite{helgaker99}.

In a typical NMR experiment, the measured quantities are the Larmor frequencies of the compound of interest, $\nu$, and that of a reference compound, $\nu_{\text{ref}}$. These define the chemical shift, which is equivalent to the difference of chemical shieldings:
\begin{equation}
  \label{eq:shift_def}
  \delta_A = \frac{\nu - \nu_{\text{ref}}}{\nu_{\text{ref}}} \simeq \sigma_{\text{ref}} - \sigma_A.
\end{equation}
Thus, two separate calculations (one for the reference and one for the compound of interest) are generally required when comparing computational predictions to experimentally obtained chemical shifts.

The magnetizability tensor is defined as the proportionality between an external magnetic field and the induced magnetic moment of the electronic subsystem of a molecule:
\begin{equation}
  \overleftrightarrow\xi = \bm m_{\text{ind}}(\bm B)/\bm B .
\end{equation}
  Like shieldings, it may be identified with a second energy derivative:
\begin{equation}
  \label{eq:xi_def}
  \overleftrightarrow\xi = - \left. \frac{\partial ^2E}{\partial \bm B^2} \right|_{\bm B = 0}.
\end{equation}
For a closed shell molecule, it describes the lowest order response of the electronic system to an external field. A related macroscopic property, the magnetic susceptibility $\overleftrightarrow\chi_M$, is connected to the magnetizability via the relation
\begin{equation}
  \label{eq:chi_mac}
  \overleftrightarrow\chi_M = \mu_0 \tilde n \overleftrightarrow\xi,
\end{equation}
where $\tilde n$ is the density of molecules (number of induced magnetic dipoles per unit volume) and $\mu_0$ is the vacuum permeability. Finite magnetic susceptibility of the sample can distort the uniformity of the external magnetic field, which can affect the resolution and peak positions of the NMR spectrum (in particular, it is partly responsible for the residual dipolar couplings) \cite{vanderhart07}.

The indirect spin-spin coupling tensor, $\overleftrightarrow J_{AB}$, describes the interaction between two nuclear spins, and it is mediated by the surrounding electron cloud. In other words, $\overleftrightarrow J_{AB}$ is defined as the difference between the total spin-spin coupling tensor and the classical dipole-dipole tensor. It may be evaluated as
\begin{equation}
  \label{eq:J_def}
  \begin{split}
    \overleftrightarrow J_{AB} =& \frac{1}{h} \left. \frac{\partial ^2 E}{\partial \bm I_A\partial \bm I_B} \right|_{\bm I_A=\bm I_B=0} \\
    =& h\frac{\gamma_A}{2\pi}\frac{\gamma_B}{2\pi} \left. \frac{\partial ^2 E}{\partial \bm\mu_A\partial \bm\mu_B} \right|_{\bm\mu_A=\bm\mu_B=0},
\end{split}
\end{equation}
where $\bm I_X$, $\gamma_X$, and $\bm\mu_X$ are the spin, gyromagnetic ratio, and the magnetic moment of nucleus $X$.

\subsection{Gauge-including atomic orbital shieldings and magnetizabilities}

Our starting point for taking the energy derivatives \eqref{eq:shield_def} and \eqref{eq:xi_def} is the
total energy expression of
nonrelativistic Kohn-Sham DFT \cite{hohenberg64},\cite{kohn65}:
\begin{equation}
  \label{eq:KS-functional}
  \begin{split}
    E[n(\bm r)] =& T_s[n(\bm r)] + \int V_{\text{ext}}(\bm r) n(\bm r) \,\mathrm{d}\bm r \\
    &+ \frac{1}{2}\int \frac{n(\bm r)n(\bm r')}{|\bm r - \bm r'|}\mathrm{d}\bm r\mathrm{d}\bm r' \\
    &+ E_\text{xc}[n(\bm r)] + E_{\text{nuc-nuc}}.
  \end{split}
\end{equation}
$n(\bm r) = n_\uparrow(\bm r) + n_\downarrow(\bm r) $ is the electron density, derived from the Kohn-Sham effective single-particle orbitals $\psi_{i,\sigma}$ in both spin channels $\sigma$ ($\sigma$ = $\uparrow$ or $\downarrow$), and $T_s[n(\bm r)]$ is the associated single-particle kinetic energy. $V_{\text{ext}}$ is the electron-nucleus potential, $E_{\text{nuc-nuc}}$ is the internuclear repulsion, and $E_\text{xc}[n(\bm r)]$ is the exchange-correlation energy. The spin densities are
\begin{equation}
  n_\sigma(\bm r) = \sum_i^{\text{occ}} |\psi_{\sigma,i}(\bm r)|^2 .
\end{equation}
(Our current implementation assumes integer occupation of the orbitals, i.e., fractional orbital occupations such as in metallic systems are not yet supported.)
The associated single-particle Hamiltonian to construct the orbitals and thus the density is
\begin{equation}
  \label{eq:H0}
  H_0 = \frac{1}{2}\bm p^2 + V_{\text{ext}} + V_{\text{H}} + V_{\text{xc}},
\end{equation}
where $\bm p = -i\bm\nabla$ is the momentum operator, $V_{\text{H}}$ is the Hartree potential, and $V_{\text{xc}}$ is the exchange-correlation (xc) potential,
\begin{equation}
V_{\text{xc}} = \frac{\delta}{\delta n} E_\text{xc}[n(\bm r)] .
\end{equation}
In semilocal DFT, the exchange-correlation energy has the form
\begin{equation}
  E_{\text{xc}}[n] = \int n(\bm r) \epsilon_{\text{xc}}[n](\bm r) \,\mathrm{d}\bm r,
\end{equation}
where the exchange-correlation energy density $\epsilon_{\text{xc}}[n](\bm r)$ is an explicit function of the density, its gradients, and/or other local quantities derived from the Kohn-Sham orbitals (and thus, effectively, a function of $\bm r$). Specifically, in the local (spin) density approximation, L(S)DA,
\begin{equation}
  \label{eq:lda}
  \epsilon_{\text{xc}}[n](\bm r) = \epsilon^\text{LSDA}_{\text{xc}}(n_\uparrow(\bm r),n_\downarrow(\bm r)) .
\end{equation}
In a generalized gradient approximation (GGA),
\begin{equation}
  \epsilon_{\text{xc}}[n](\bm r) = \epsilon_{\text{xc}}^{\text{GGA}}(n_\uparrow,n_\downarrow,|\nabla n_\uparrow|,|\nabla n_\downarrow|) .
\end{equation}

In practice, the magnetic perturbations are introduced via Eq.~\eqref{eq:H0} and the minimal coupling substitution,
\begin{equation}
  \label{eq:min_coupling}
  \bm p \rightarrow \bm p + \bm A,
\end{equation}
where $\bm A$ is the magnetic vector potential. This yields the full Hamiltonian
\begin{equation}
  \label{eq:H}
  H = H_0 + \bm p\bm A + \frac{1}{2}\bm A^2.
\end{equation}
In the presence of an external magnetic field and an NMR active nucleus, the vector potential may be taken to assume the form
\begin{equation}
  \label{eq:A}
  \bm A = \frac{1}{2}\bm B\times(\bm r-\bm R_0) + \alpha^2\frac{\bm\mu_A\times \bm r_A}{r_A^3},
\end{equation}
where $\bm r$ is the position vector, $\bm R_0$ is the gauge origin, $\bm r_A = \bm r - \bm R_A$ is the position vector relative to the nucleus $A$, and $\alpha$ is the fine structure constant.

The gauge origin is arbitrary and in principle should not influence any observables. In practical calculations, with a finite basis set, some magnetic properties show a strong dependence on the gauge origin and exhibit poor convergence unless a gauge-invariant formalism is used. This trend arises from the two terms making up the shielding tensor, the paramagnetic and diamagnetic part, which converge at different rates because the former contains a sum over unoccupied states while the latter does not \cite{gregor99}. The same logic holds for the magnetizability. We follow the standard approach of gauge-including atomic orbitals (GIAOs) \cite{ditchfield72}, where each basis function $\phi_n$ is equipped with a phase factor of the form
\begin{equation}
  \label{eq:giao}
  \phi_n^{\text{GIAO}}(\bm r) = e^{-i\bm A_n\bm r} \phi_n(\bm r) = e^{-\frac{i}{2} (\bm R_n\times\bm r)\bm B} \phi_n(\bm r),
\end{equation}
where $\bm A_n = \frac{1}{2} \bm B\times (\bm r - \bm R_n)$ is the vector potential with the gauge origin shifted to the origin of the basis function $n$, which is at $\bm R_n$. GIAOs are approximate solutions to a one-electron system that has been perturbed by an external field, which motivates their use as basis functions in magnetic response calculations. The matrix elements of the Kohn-Sham Hamiltonian then become
\begin{equation}
  \label{eq:giao_el}
  \begin{split}
    &\braket{\phi_m^{\text{GIAO}} | H | \phi_n^{\text{GIAO}}} \\
    &= \braket{\phi_m | e^{\frac{i}{2}(\bm R_m\times\bm r)\bm B} H e^{-\frac{i}{2}(\bm R_n\times\bm r)\bm B} | \phi_n} \\
    &= \braket{\phi_m | e^{\frac{i}{2}(\bm R_{mn}\times\bm r)\bm B} \left[ H_0 + \bm p\bm A_n + \frac{1}{2}\bm A_n^2 \right] | \phi_n},
  \end{split}
\end{equation}
where $\bm R_{mn} = \bm R_m - \bm R_n$ and where $\bm A$ has been replaced by $\bm A_n$, i.e., the dependence on the gauge origin has been eliminated.

With GIAOs and rewriting Eq.~\eqref{eq:KS-functional} in terms of the eigenvalues of $H$, $\braket{\psi_m^{\text{GIAO}} | H | \psi_m^{\text{GIAO}}}$, the total energy in \eqref{eq:shield_def} and \eqref{eq:xi_def} is
\begin{equation}
  \begin{split}
    E =& \sum_m^{\text{occ}} \braket{\psi_m^{\text{GIAO}} | H | \psi_m^{\text{GIAO}}} \\
    &- \frac{1}{2}\int \frac{n^{\text{GIAO}}(\bm r)n^{\text{GIAO}}(\bm r')}{|\bm r - \bm r'|}\mathrm{d}\bm r\mathrm{d}\bm r' \\
    &- \int n^{\text{GIAO}}(\bm r)v_{\text{xc}}(n^{\text{GIAO}}(\bm r))\mathrm{d}\bm r \\
    &+ E_{\text{xc}}[n^{\text{GIAO}}] + E_{\text{nuc-nuc}},
  \end{split}
\end{equation}
where
\begin{equation}
  n^{\text{GIAO}}(\bm r) = \sum_n^{\text{occ}} |\psi_n^{\text{GIAO}}(\bm r)|^2
\end{equation}
is the total density,
\begin{equation}
  \psi_n^{\text{GIAO}}(\bm r) = \sum_i C_{in}\phi_i^{\text{GIAO}}(\bm r)
\end{equation}
are the molecular orbitals and $\bm C$ the molecular orbital coefficients.

If the basis functions were independent of the perturbation, taking the total energy derivatives would be fairly straightforward. With GIAOs, however, we must use the general equation for mixed second derivatives of the total energy \cite{pople72}. Using the notation
\begin{equation}
  f^{\bm M} \equiv \frac{\partial  f}{\partial \bm M}; \quad f^{\bm M\bm N} \equiv \frac{\partial ^2 f}{\partial \bm M\partial \bm N}
\end{equation}
for first and second derivatives, the second energy derivative is
\begin{equation}
  \label{eq:E2_general}
  \begin{split}
    E^{\bm M\bm N} =& \sum_m^{\text{occ}} \sum_{ij} C_{im}H_{ij}^{\bm M\bm N}C_{jm}
    \\
    &+ \sum_m^{\text{occ}} \sum_{ij} 2\mathrm{Re} C_{im}^{\bm N} \left( H_{ij}^{\bm M} - S_{ij}^{\bm M}\epsilon_m \right) C_{jm} \\
    &- \sum_m^{\text{occ}} \sum_{ij} C_{im} S_{ij}^{\bm M\bm N} C_{jm} \epsilon_m \\
    &- \sum_{mn}^{\text{occ}} \sum_{ij} C_{im} S_{ij}^{\bm M} C_{jm} \epsilon_{\bm N mn}.
  \end{split}
\end{equation}
$\bm C$ are the unperturbed molecular orbital coefficients,
\begin{equation}
  \psi_n(\bm r) = \sum_i C_{in} \phi_i(\bm r),
\end{equation}
and $\bm C^{\bm N}$ are the first-order orbital coefficients,
\begin{equation}
  \psi_n^{\bm N}(\bm r) = \sum_i C_{in}^{\bm N} \phi_i(\bm r).
\end{equation}
$\psi_n^{\bm N}$ are the first-order orbitals, $H_{ij}^{\bm M}$ and $H_{ij}^{\bm M\bm N}$ are the first and second derivatives of the Hamiltonian matrix elements, $S_{ij}^{\bm M}$ and $S_{ij}^{\bm M\bm N}$ are the first and second derivatives of the overlap matrix elements, $\epsilon_m$ are the unperturbed eigenvalues, and
\begin{equation}
  \epsilon_{\bm M mn} = \braket{\psi_m | H^{\bm M} | \psi_n} - \frac{\epsilon_m + \epsilon_n}{2} \braket{\psi_m | S^{\bm M} | \psi_n}.
\end{equation}
In the case of GIAOs, derivatives of the overlap matrix elements are
\begin{equation}
  \begin{split}
    S_{ij}^{\bm B} =& \frac{i}{2} \braket{\phi_i| \bm R_{ij}\times\bm r |\phi_j}, \\
    S_{ij}^{\bm B\bm B} =& -\frac{1}{4} \braket{\phi_i| (\bm R_{ij}\times\bm r)(\bm R_{ij}\times\bm r)^T |\phi_j}.
  \end{split}
\end{equation}
The first-order orbital coefficients are obtained from density functional perturbation theory (DFPT) \cite{schreckenbach95},\cite{dupuis01}, where the basic \textit{ansatz} is that quantities such as the electron density may be written as a perturbation series in terms of a perturbation $\lambda$:
\begin{equation}
  n = n^{(0)} + \lambda n^{(1)} + \lambda^2 n^{(2)} + \ldots
\end{equation}
We define the perturbed molecular coefficients $\bm C^{\bm M}$ as
\begin{equation}
  \label{eq:wf1_def}
  C_{in}^{\bm M} = \sum_k C_{ik}U_{kn}
\end{equation}
in terms of the unperturbed coefficients $\bm C$ and the $\bm U$-matrices, where
\begin{equation}
  U_{mn} = -\frac{1}{2} \sum_{ij} C_{im} S_{ij}^{\bm M} C_{jn}
\end{equation}
if $mn$ is an occupied-occupied or unoccupied-unoccupied pair, and
\begin{equation}
  U_{mn} = \sum_{ij} \frac{C_{im} H_{ij}^{\bm M} C_{jn} -  C_{im} S_{ij}^{\bm M} C_{jn} \epsilon_n}{\epsilon_n - \epsilon_m}
\end{equation}
otherwise.

In the case of shieldings, the perturbations $\bm M$ and $\bm N$ are taken to be the nuclear magnetic moment $\bm\mu_A$ and the magnetic field $\bm B$. Because the basis functions do not explicitly depend on the magnetic moment, $\bm S^{\bm\mu_A} = \bm S^{\bm B\bm\mu_A} = 0$, the perturbed overlap matrices vanish in Eq.~\eqref{eq:E2_general}. In the case of magnetizability, both $\bm M$ and $\bm N$ correspond to the $\bm B$ perturbation and nothing vanishes in Eq.~\eqref{eq:E2_general}.

Using Eq.~\eqref{eq:giao_el} for the Hamiltonian matrix elements, with the vector potential given by Eq.~\eqref{eq:A}, we can take the derivatives necessary for evaluating the second total energy derivative, Eq.~\eqref{eq:E2_general}:
\begin{equation}
  \label{eq:terms}
  \begin{split}
    H_{mn}^{\bm B} =& \frac{i}{2} \Braket{\phi_m | (\bm R_{mn}\times\bm r) H_0 - \bm r_n\times\bm\nabla | \phi_n} \\
    & + \braket{\phi_m | V_{\text{GGA}}^{\bm B} | \phi_n}, \\
    H_{mn}^{\bm B\bm\mu_A} =& \frac{\alpha^2}{2} \Braket{\phi_m | \frac{(\bm R_{mn}\times\bm r)(\bm r_A\times\bm\nabla)^T}{r_A^3} | \phi_n} \\
    &+ \frac{\alpha^2}{2} \Braket{\phi_m | \frac{\bm r_A\bm r_n - \bm r_A\bm r_n^T}{r_A^3} | \phi_n}, \\
    H_{mn}^{\bm\mu_A} =&  -i\alpha^2 \Braket{\phi_m | \frac{\bm r_A\times \bm\nabla}{r_A^3} | \phi_n}, \\
    H_{mn}^{\bm B\bm B} =& \frac{1}{4} \Braket{\phi_m | 2(\bm R_{mn}\times\bm r) (\bm r_n \times \bm\nabla)^T | \phi_n} \\
    &+ \frac{1}{4} \Braket{\phi_m | r_n^2 - \bm r_n\bm r_n^T | \phi_n} \\
    &- \frac{1}{4} \Braket{\phi_m | (\bm R_{mn}\times\bm r)(\bm R_{mn}\times\bm r)^T H_{\text{LDA}} | \phi_n} \\
    &+ \Braket{\phi_m | V_{\text{GGA}}^{\bm B\bm B} | \phi_n}.
  \end{split}
\end{equation}
In Eq.~\eqref{eq:E2_general}, terms containing second-order operators such as $H^{\bm B\bm\mu_A}$ or $H^{\bm B\bm B}$ are usually called the diamagnetic contribution of the magnetic property. These require computing expectation values between the unperturbed orbitals. The rest is called the paramagnetic contribution, which requires computing expectation values between unperturbed and perturbed orbitals. The paramagnetic and diamagnetic contributions are sometimes defined differently by various authors \cite{ruud94},\cite{schreckenbach95},\cite{skachkov10}. This distinction has no physical consequences as long as their sum is invariant. The matrix elements as shown in Eqs.~\eqref{eq:terms} are grouped simply to make the computation as efficient as possible.

With a GGA functional, the first $\bm B$ derivative of the gradient part of the xc-functional is \cite{helgaker00}
\begin{multline}
  \braket{\phi_m | V_{\text{GGA},\alpha}^{\bm B} | \phi_n} = \frac{i}{2} \int \,\mathrm{d}\bm r\, \bm\nabla (\phi_m\phi_n) \\
  \times (\bm R_{mn}\times\bm r) \left( 2\frac{\partial E_{\text{xc}}}{\partial \gamma_{\alpha\alpha}}\bm\nabla n_\alpha + \frac{\partial E_{\text{xc}}}{\partial \gamma_{\alpha\beta}}\bm\nabla n_\beta \right) \\
  + \frac{i}{2} \int \,\mathrm{d}\bm r\, \phi_m\phi_n \\
  \times \left[ 2\frac{\partial E_{\text{xc}}}{\partial \gamma_{\alpha\alpha}} (\bm R_{mn}\times\bm\nabla n_\alpha) + \frac{\partial E_{\text{xc}}}{\partial \gamma_{\alpha\beta}} (\bm R_{mn}\times\bm\nabla n_\beta) \right].
\end{multline}
The second $\bm B$ derivative is
\begin{multline}
  \braket{\phi_m | V_{\text{GGA},\alpha}^{\bm B\bm B} | \phi_n}  = -\frac{1}{2} \int \,\mathrm{d}\bm r\, \phi_m\phi_n \\
  \times \bm R_{mn} \times \left( 2\frac{\partial E_{\text{xc}}}{\partial \gamma_{\alpha\alpha}} \bm\nabla n_\alpha + \frac{\partial E_{\text{xc}}}{\partial \gamma_{\alpha\beta}} \bm\nabla n_\beta \right) (\bm R_{mn}\times\bm r)^T \\
  -\frac{1}{4} \int \,\mathrm{d}\bm r\, \bm\nabla (\phi_m\phi_n) \\
  \times (\bm R_{mn}\times\bm r)(\bm R_{mn}\times\bm r)^T \left( 2\frac{\partial E_{\text{xc}}}{\partial \gamma_{\alpha\alpha}}\bm\nabla n_\alpha + \frac{\partial E_{\text{xc}}}{\partial \gamma_{\alpha\beta}}\bm\nabla n_\beta \right)
\end{multline}
for spin channel $\alpha$ (for spin population $\uparrow$ or $\downarrow$). In the LDA, the $V_{\text{GGA}}^{\bm B}$ and $V_{\text{GGA}}^{\bm B\bm B}$ terms vanish.

With a hybrid density functional, i.e. if a portion of exact exchange were included in the exchange-correlation energy, the GIAO formalism would lead to additional terms \cite{helgaker91}. This would require further modifications to the resolution of identity technique that is used to evaluate the 4-center integrals in FHI-aims \cite{ren12},\cite{ihrig15}. The complication is that with GIAOs, the 4-center integrals would have a contribution from the position vector in the integrand. In this study, Hartree-Fock and hybrid functionals are therefore not considered.

In contrast to \textit{J}-couplings (see below), which involve real-valued perturbations, the perturbations considered for shieldings and magnetizabilities are purely imaginary. For imaginary perturbations there is no first-order density and no first-order response from the Hartree or exchange-correlation potentials that would otherwise need to be computed self-consistently \cite{sychrovsky00}. Because the perturbations are imaginary, there is also no first-order density and first-order Hartree potential even for open-shell systems. Thus, our implementation of shieldings and magnetizabilities covers both closed and open-shell systems.

\subsection{\textit{J}-couplings}

\textit{J}-couplings are derived within the same formalism as above by taking the $M$ and $N$ perturbations as magnetic moments $\bm\mu_A$ and $\bm\mu_B$. The vector potential now only includes contributions from the two magnetic moments:
\begin{equation}
  \label{eq:J_A}
  \bm A = \alpha^2 \left( \frac{\bm\mu_A\times \bm r_A}{r_A^3} + \frac{\bm\mu_B\times \bm r_B}{r_B^3} \right).
\end{equation}
Because the GIAOs need not be used for \textit{J}-couplings, all first and higher order derivatives of the overlap matrix vanish. It is then straightforward to take the first and second derivatives that enter Eq.~\eqref{eq:E2_general} \cite{sychrovsky00}:
\begin{align}
  \label{eq:FC}
    H_{mn}^{\text{FC}\bm\mu_A} =& \frac{8\pi\alpha^2}{3} \phi_m(\bm R_A) \hat S \phi_n(\bm R_A), \\
    H_{mn}^{\text{PSO}\bm\mu_A} =&  -i\alpha^2 \Braket{\phi_m | \frac{\bm r_A\times \bm\nabla}{r_A^3} | \phi_n}, \\
    H_{mn}^{\text{SD}\bm\mu_A} =&  \alpha^2 \Braket{\phi_m | \frac{3(\hat S\bm r_A) \bm r_A}{r_A^5} - \frac{\hat S}{r_A^3} | \phi_n}, \\
    H_{mn}^{\bm\mu_A,\bm\mu_B} =& \alpha^4 \Braket{\phi_m | \frac{\bm r_A\bm r_B - \bm r_B\bm r_A^T}{r_A^3r_B^3} | \phi_n},
\end{align}
where $\hat S$ is the spin operator. The four terms that make up \textit{J}-couplings are the following. The Fermi contact (FC) term, which is evaluated at a single point at the nucleus of interest, generally makes up the bulk of the \textit{J}-coupling. The other terms, which represent less localized coupling mechanisms, are the paramagnetic spin-orbit (PSO), which usually comes after FC in terms of magnitude, and the spin-dipole (SD) and diamagnetic spin-orbit terms. The SD term, in particular, is computationally most demanding because of the scalar product $\hat S\bm r_A$. It is often very small, but can be considerable in some systems. Because it is difficult to say \textit{a priori} in which systems it plays a role \cite{sychrovsky00}, it can never be simply omitted.

In contrast to the imaginary operators used in the calculation of $\overleftrightarrow\sigma_A$ and $\overleftrightarrow\xi$, the FC and the SD operators represent real-valued perturbations, which give rise to a nonzero first-order electron density. This produces a first-order response of the xc-potential, which itself is a functional of the first-order density
\begin{equation}
  \label{eq:rho1}
  n^{\bm\mu_A}(\bm r) = \sum_m^{\text{occ}}\sum_{ij} \phi_i(\bm r) \left[ C_{im}C_{jm}^{\bm\mu_A*} + C_{im}^{\bm\mu_A}C_{jm} \right] \phi_j(\bm r).
\end{equation}
The $\bm U$ matrix, defined by Eq.~\eqref{eq:wf1_def}, is now given by
\begin{equation}
  \label{eq:U_j}
  U_{mn} = \sum_{ij} \frac{C_{im} H_{ij}^{\bm\mu_A} C_{jn} + C_{im} V_{xc;ij}^{\bm\mu_A}[n^{\bm\mu_A}] C_{jn}}{\epsilon_n - \epsilon_m},
\end{equation}
where the first-order xc-potential $V_{xc;ij}^{\bm\mu_A}$ depends on the first-order density. The xc-kernel matrix elements needed for $V_{xc;ij}^{\bm\mu_A}$ are computed using procedures from the LibXC library \cite{marques12} and the full expressions are shown in the Appendix. Thus, the DFPT cycle, which consists of Eqs.~\eqref{eq:wf1_def}, \eqref{eq:U_j}, and \eqref{eq:rho1}, must now be performed self-consistently.

In case of an open-shell system, Eq.~\eqref{eq:U_j} contains an additional term from the first-order Hartree potential,
\begin{equation}
  \begin{split}
    & U_{mn} \rightarrow U_{mn} + \sum_{ij} \frac{C_{im} V_{H;ij}^{\bm\mu_A}[n^{\bm\mu_A}] C_{jn}}{\epsilon_n - \epsilon_m}, \\
    & V_{H;ij}^{\bm\mu_A}[n^{\bm\mu_A}] = \int \frac{n^{\bm\mu_A}(\bm r')\phi_i(\bm r)\phi_j(\bm r)}{\bm r-\bm r'} \,\mathrm{d}\bm r,
  \end{split}
\end{equation}
which is also included in our implementation.

\subsection{Numeric atom-centered orbitals}
\label{sec:naos}

A detailed overview of numeric atom-centered orbitals, as implemented in FHI-aims, is available in Ref.~\cite{blum09}. The magnetic response properties can be implemented into that framework without major obstacles.

All integration routines are based on the partitioning of unity \cite{delley90},\cite{becke88}. Each integral is divided into atom-centered pieces according to
\begin{multline}
  \int \phi_m(\bm r) \hat{\mathcal O} \phi_n(\bm r) \,\mathrm{d}\bm r \\
  = \sum_A \int p_A(\bm r) \phi_m(\bm r) \hat{\mathcal O} \phi_n(\bm r) \,\mathrm{d}\bm r,
\end{multline}
where $\hat{\mathcal O}$ is a general operator and the sum is over the atom-centered partition functions $p_A(\bm r)$, defined such that $\sum p_A(\bm r) = 1$ at all grid points. The partition functions determine the weights of grid points in the single-atom integrands, which are all integrated on their own atom-centered grids. The grid points are distributed on spherical shells around each nucleus at certain radii $r(s)$ (in the simplest case, $s=1,\ldots,N_r$). The radial grids $r(s)$ are determined by three parameters. Two parameters define a basic radial grid, given by the radius of the outermost shell, $r_{\text{outer}}$, and the number of points, $N_r$ \cite{baker94}:
\begin{equation}
  \label{eq:radial_grid}
  r(s) = r_{\text{outer}} \frac{\log\{1-[s/(N_r+1)]^2\}}{\log\{1-[N_r/(N_r+1)]^2\}}.
\end{equation}
The density of this basic grid can be increased by a third parameter (\texttt{radial\_multiplier}), placing additional points at integer fractions of the original grid, e.g., at $s = 1/2, \ldots, N_r+1/2$ for \texttt{radial\_multiplier}=2, and similar for other integer values (see appendix of Ref.~\cite{zhang13} for an illustration). The angular distribution of the grid points distributed on each radial shell is determined by a modified Lebedev algorithm \cite{delley96}. The number of angular grid points per shell increases with increasing radius $r(s)$.

Furthermore, with confining potential techniques \cite{koepernik99},\cite{horsfield97},\cite{sankey89},\cite{eschrig78},\cite{eschrig89},\cite{delley00},\cite{junquera01},\cite{ozaki04}, the radial functions can be chosen to be strictly localized within a given cutoff radius \cite{blum09}, which allows the otherwise most expensive parts of the calculations --- integrals, charge density updates, and, for hybrid density functionals, the exchange operator --- to scale linearly with system size ($\mathcal O(N)$ scaling). This is relevant for NMR, which often deals with large-scale systems such as biomolecules \cite{edwards14}. In FHI-aims, this confinement radius is a configurable parameter. By default, the extent of a radial function is limited to 5~Å (6~Å) for ``light'' (``tight'') settings for most elements.

With NAOs, it is natural to parallelize the integrations over grid points. In FHI-aims, efficient parallelization is achieved by dividing the global grid into localized subsets (``batches'') of grid points
\begin{equation}
  \begin{split}
    h_{mn} =& \sum_{\text{batch}}^{N_\text{batches}} h_{mn}^{\text{batch}}, \\
    h_{mn}^{\text{batch}} =& \sum_{\bm r \text{ in batch}} w(\bm r)\phi_m(\bm r) \hat h \phi_n(\bm r),
\end{split}
\end{equation}
as detailed in Ref.~\cite{havu09}. $w(\bm r)$ are the integration weights, which include both radial and angular integration weights and the partition functions \cite{blum09}. Generally, each batch contains a few hundred grid points. Because all basis functions can be constructed so that they go to zero beyond a cutoff radius, only a certain number of nonzero functions needs to be considered within each batch. As the system size is increased, the number of basis functions relevant to each batch reaches a plateau. This leads to $\mathcal O(N)$ scaling of all integrations for sufficiently extended systems, where $N$ is some measure of the system size such as the number of atoms.

Within each batch, the matrix elements $h_{mn}^{\text{batch}}$ are evaluated by first applying the given operator to a basis function on each grid point, $\hat h \phi_n(\bm r)$, and then multiplying with $\phi_m(\bm r)$ using a BLAS (basic linear algebra subroutines) level 3 routine for all $mn$. BLAS level 3 are matrix-matrix multiplication type operations and thus very efficient. A complication arises with the GIAO integrals due to the $\bm R_{mn}$ factors, which make the operators $\hat h^{\text{batch}}$ [$\hat H^{\bm M}$ and $\hat H^{\bm M\bm N}$ in Eq.\eqref{eq:E2_general}] dependent on both sides of the expectation value. Efficient matrix-matrix multiplications can still be performed if the $\bm R_{mn}$ factors are taken out of the integrals and later multiplied with the result elementwise. For integrals such as the first-order overlap matrix,
\begin{equation}
  S_{mn}^{\bm B} = \frac{i}{2} \bm R_{mn}\times \braket{\phi_m| \bm r |\phi_n},
\end{equation}
the working memory usage for the matrix elements is increased by approximately three times compared to the case where GIAOs are not used due to the requirement of keeping all directions of the position operator in memory. Integrals such as the second-order overlap matrix, which contain two cross products of $\bm R_{mn}$, may be rewritten as
\begin{equation}
  S_{mn}^{\bm B\bm B} = \frac{1}{4} \bm R_{mn}\times \braket{\phi_m| \bm r\bm r^T |\phi_n} \times \bm R_{mn},
\end{equation}
and require approximately 9 times as much memory, because all components of the $\bm r\bm r^T$ tensor must be available before taking the cross products. This is seen by explicitly writing out, e.g., the diagonal components of a tensor $\overleftrightarrow X = \bm R \times \overleftrightarrow H \times \bm R$:
\begin{equation}
  \begin{split}
    X_{11} =& R_2 (H_{32} R_3-H_{33} R_2) - R_3 (H_{22} R_3-H_{23} R_2), \\
    X_{22} =& R_3 (H_{13} R_1-H_{11} R_3) - R_1 (H_{33} R_1-H_{31} R_3), \\
    X_{33} =& R_1 (H_{21} R_2-H_{22} R_1) - R_2 (H_{11} R_2-H_{12} R_1).
  \end{split}
\end{equation}
Other than a larger prefactor, memory usage still scales as $\mathcal O(N)$ with system size and can easily be controlled with the number of compute tasks on sufficiently large parallel machines.

\section{Results and discussion}
\label{sec:results}

\subsection{Test cases and numerical choices}

In section \ref{sec:results}, we present basis set convergence studies for the different NMR parameters --- chemical shieldings (and shifts), magnetizabilities, and \textit{J}-couplings. We compare the isotropic value or trace of all tensorial quantities. All calculations were performed in the PBE parameterization of exchange and correlation. We consider the FHI-aims-09, NAO-VCC-$n$Z, NAO-J-$n$, and the GTO-based (aug-)cc-pV$n$Z, (aug-)cc-pCV$n$Z, (aug-)pc-$n$, (aug-)pcS-$n$, and pcJ-$n$ basis sets. Since GTOs are special cases of the generic form of Eq.~\eqref{eq:nao_def}, they are readily available in FHI-aims and can be compared to the NAOs using the exact same numerical framework.

\begin{figure}
  \centering
  \includegraphics{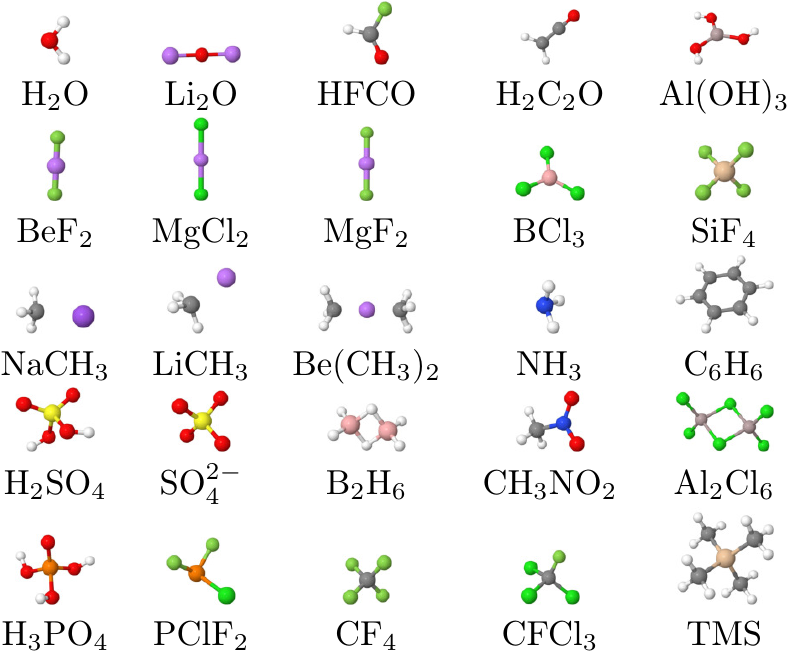}
  \caption{Test set of molecules for the magnetic response calculations. The coloring scheme is as follows: H: white, C: grey, N: blue, O: red, F: dark green, Li/Be/Na/Mg: purple, B: pink, Al: light brown, Si: tan, P: orange, S: yellow, Cl: bright green. Molecules that were used only for shieldings and magnetizabilities but not for \textit{J}-couplings are Be(CH$_3$)$_2$, BeF$_2$, Li$_2$O, LiCH$_3$, MgCl$_2$, MgF$_2$, and NaCH$_3$. TMS stands for tetramethylsilane. Geometries used for all calculations are found in the Supporting Information.}
  \label{fig:molecules}
\end{figure}

For the basis set convergence study, we use a test set of 25 molecules, shown in Fig.~\ref{fig:molecules}.
The test set of molecules covers a wide range of different chemical bonding types and contains most elements of H through Cl in the periodic table. Before calculating the NMR parameters, all geometries were optimized using the PBE functional \cite{perdew96} with a van der Waals correction \cite{tkatchenko09} and the FHI-aims-09 tier 2 basis sets. We calculated an overall number of 57 shieldings, 46 \textit{J}-couplings, and 24 magnetizabilities for different elements (see SI, Figs.~S1--S93 and Tables~S1--S75, for details about geometries and convergence studies for the magnetic properties). The implementation was also tested for open-shell systems. Exemplary results for shieldings and J-couplings of NO$_2$ can be found in the Supporting Information (Figs.~S113 and S114).

In addition to the set of small molecules, we test the applicability of our implementation to three larger systems:

First, the 4-(2-Phenylethynyl)pyridine (PEP) molecule, shown as an inset in the relevant Figure, Fig.~\ref{fig:PPA_shield}.

\begin{figure}
  \centering
  \includegraphics[scale=0.3]{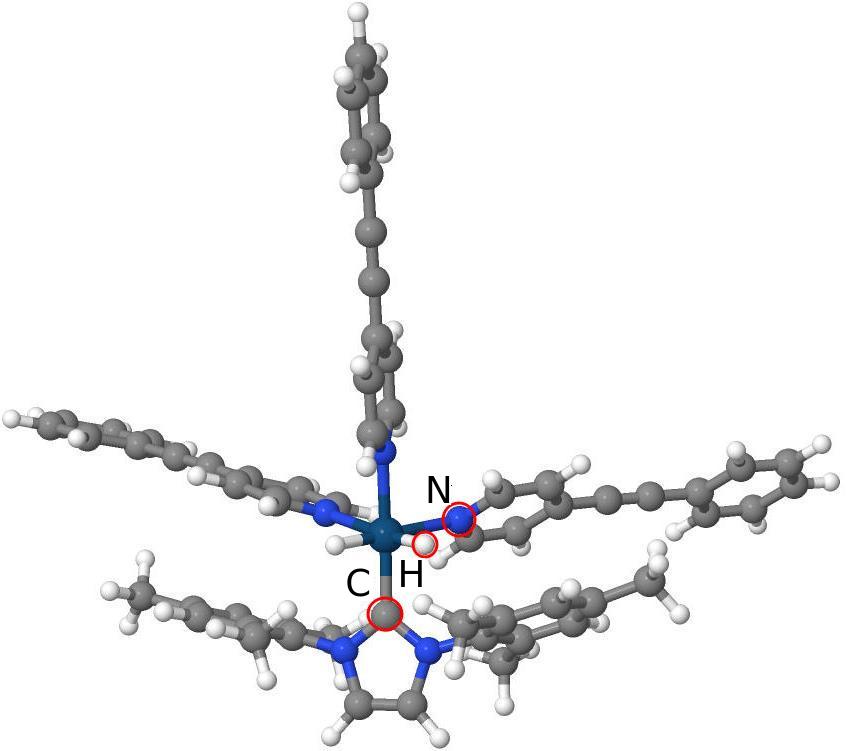}
  \caption{PTC molecule. Shown are sites for which the chemical shieldings were calculated.}
  \label{fig:ir}
\end{figure}

Second, the transition metal complex [IrH$_2$(IMes)(PEP)$_3$]$^+$ (IMes = 1,3-bis(2,4,6-trimethylphenyl)imidazole-2-ylidene), shown in Fig.~\ref{fig:ir}, with sites for which shieldings are calculated marked in red.
This molecule is one example of a polarization transfer catalyst (PTC) species in \textit{para}-hydrogen based hyperpolarization \cite{theis16} and is thus of practical importance \cite{adams09},\cite{cowley11}. The compound contains 119 atoms, including one heavy element (Ir) necessitating use of the scalar-relativistic atomic ZORA approximation as defined in Eqs.~(55) and (56) of Ref.~\cite{blum09}. The resulting relativistic density and orbitals were used for the calculation of the magnetizability and chemical shieldings in the nonrelativistic formalism (we only consider chemical shieldings related to the light elements in the PTC, not involving the Ir atom directly).

\begin{figure}
  \centering
  \includegraphics[scale=0.4]{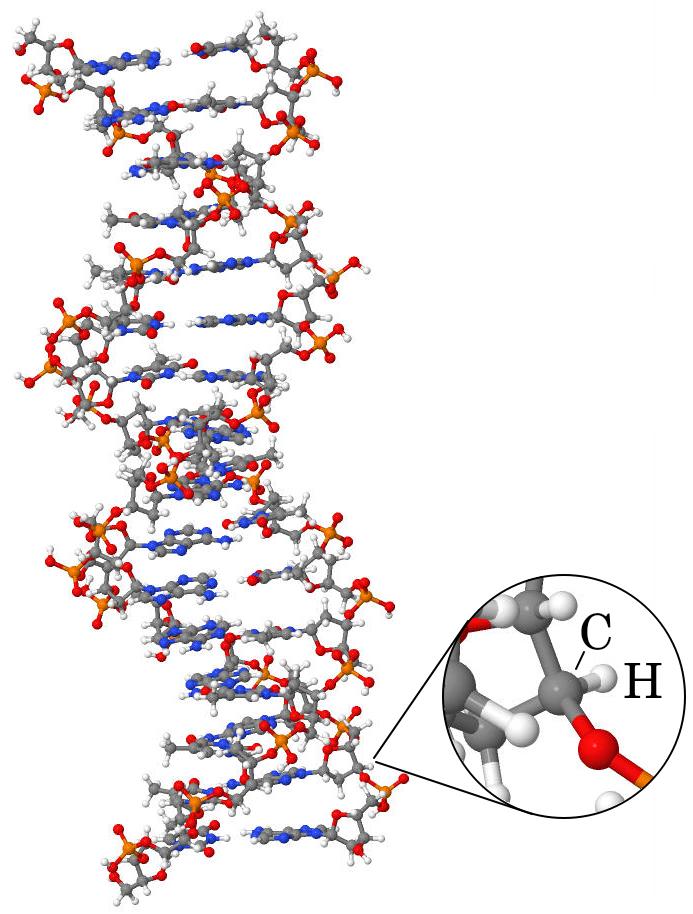}
  \caption{Geometry of the DNA segment. The shielding constant was calculated for atom 35 (C). The \textit{J}-coupling constant was calculated between atoms 29 and 35 (H and C).}
  \label{fig:dna}
\end{figure}

Third, the largest molecule in our study is a DNA segment consisting of 16 base pairs with a total of 1052 atoms (Fig.~\ref{fig:dna}).
The geometry is the same as that used by C. Ochsenfeld's group in Ref.~\cite{ochsenfeld11}.

For the calculation of shieldings, magnetizabilities and \textit{J}-couplings we first evaluated the grid settings required to obtain high-accuracy numerical results. To that end, we uniformly increased the number of radial grid shells using the \texttt{radial\_multiplier} parameter as described in Sec.~\ref{sec:naos}. In principle, it would be possible to devise a more efficient radial grid that only increases the density of grid shells near the nucleus, but we did not pursue that approach in this work.

For shieldings and magnetizabilities calculated using the FHI-aims-09 and NAO-VCC-$n$Z basis sets, we generally find the default ``tight'' settings for integration grids established in Ref.~\cite{blum09}, \texttt{radial\_multiplier}=2, to be sufficient. An exception is the NAO-VCC-5Z basis set of sulfur in the calculation of shieldings, for which \texttt{radial\_multiplier}=3 was used. The integration of tight primitive GTOs near the nucleus requires denser grids than the default grids that integrate NAOs with high accuracy \cite{zhang13}. For the correlation consistent GTO basis sets, we used radial integration grids with \texttt{radial\_multiplier}=6. Finally, for \textit{J}-couplings calculated using the NAO-J-$n$ and pcJ-$n$ basis sets, we employed \texttt{radial\_multiplier}=8. This is due to the inclusion of tight Gaussian s functions in the basis sets designed for $J$-couplings.

To facilitate a comparison of different basis sets over a set of test of compounds it is convenient to introduce the mean average error (MAE) of the calculated property $X$ ($X$ = magnetizability, chemical shielding/shift, \textit{J}-coupling) defined as
\begin{equation}
  \text{MAE}_R(B) = \sum_i^N \frac{|X_i(B)-X_i(R)|}{N},
\end{equation}
where $X_i(B)$ is the value of the property corresponding to a single calculation (e.g., shielding constant of a specific atom) obtained with basis set $B$, $X_i(R)$ is the reference value obtained with basis set $R$, and $N$ is the total number of calculations for that property (for most molecules, more than one shielding or $J$-coupling are calculated and included in the averages). The MAE describes the absolute mean deviation from reference values and thus represents a hard estimate of the basis set error. Relative errors, such as the relative position of two peaks in an NMR spectrum, could be less than what is suggested by the MAE.

In the figures given below, the MAE is plotted as a function of the basis set size. For individual molecules, the ``basis set size'' is a well defined quantity for each investigated basis set. However, when averages are taken over multiple molecules in a test set, the ``basis set size'' depends on each molecule and is no longer a uniquely defined metric. We therefore resort to the following quantity to represent the ``average basis set size'' (average number of basis functions per molecule in a test set, $N^\text{ave}$) for data that are averaged over groups of molecules:
\begin{equation}
  N^{\text{ave}} = \sum_i^{N_\text{molecules}} N^i/N_{\text{molecules}},
\end{equation}
where $N^i$ is the number of basis functions for molecule $i$ and the sum is over all molecules in the test set.

\subsection{Chemical shieldings and shifts}

\begin{figure}
  \centering
  \includegraphics[scale=1.35]{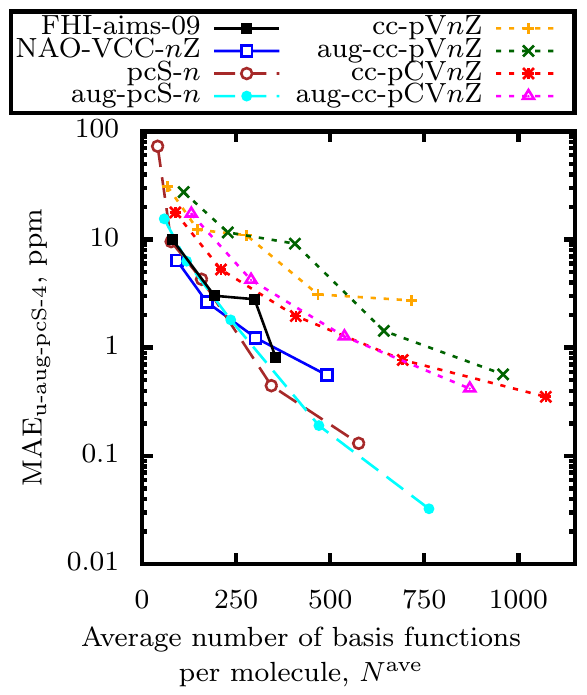}
  \caption{Mean average errors of DFT-PBE calculated shieldings for the test set of molecules shown in Fig.~\ref{fig:molecules}, using the u-aug-pcS-4 basis set as a reference. The levels shown for each basis set are the same as in Table~\ref{tab:basis_sets}.}
  \label{fig:abs_shield_err}
\end{figure}

In Fig.~\ref{fig:abs_shield_err}, we investigate the convergence of calculated chemical shieldings for the broader set of compounds introduced in Fig.~\ref{fig:molecules} as a function of basis set size and basis set type. 
The MAE of NMR shieldings is shown for different NAO and GTO basis sets, using the uncontracted aug-pcS-4 (u-aug-pcS-4) basis \cite{jensen08} as reference.

In the small averaged basis set size region (up to about 200 basis functions per molecule on average), the general-purpose FHI-aims-09 and NAO-VCC-$n$Z show similar performance as the aug-pcS-$n$ basis sets, which were optimized for shieldings. The GTO correlation consistent basis sets show a larger deviation from the reference basis sets. As the basis set size approaches the converged limit, the (aug-)cc-pV$n$Z and (aug-)cc-pCV$n$Z basis sets exhibit slower convergence with basis set size than the NAO-VCC-$n$Z or FHI-aims-09 basis sets, which remain close to the (aug)-pcS-n basis sets to below 1~ppm accuracy. For even smaller deviations, the exact MAE may reflect some bias towards the aug-pcS-$n$ basis sets, because the uncontracted aug-pcS-4 basis set from the same series serves as reference.

It is also evident that the FHI-aims-09 basis sets at the tier 3 level show a deviation from smooth convergence, while tier 2 and tier 4 on their own would otherwise show a consistent trend. This behavior reflects fluctuations in a number of the test systems at the tier 3 level, e.g., H$_2$O (Fig. S3) and NH$_3$ (Fig. S62). In these cases, it is possible to trace the fluctuation to the shape of specific basis functions in the tier 3 basis sets of O and N. Evidently, the response of the electronic system to the external perturbation is not adequately captured when these basis functions are included on their own, an issue that is alleviated by further enlarging the basis set at the tier 4 level. Overall, Fig.~\ref{fig:abs_shield_err} implies that the NAO-VCC-$n$Z basis sets are better suited for systematic convergence studies of shieldings, which is consistent with their initial construction targeting unoccupied-state sums in perturbation theory (Table~\ref{tab:basis_sets} and Ref.~\cite{zhang13}).

\begin{figure}
  \centering
  \includegraphics[scale=1.35]{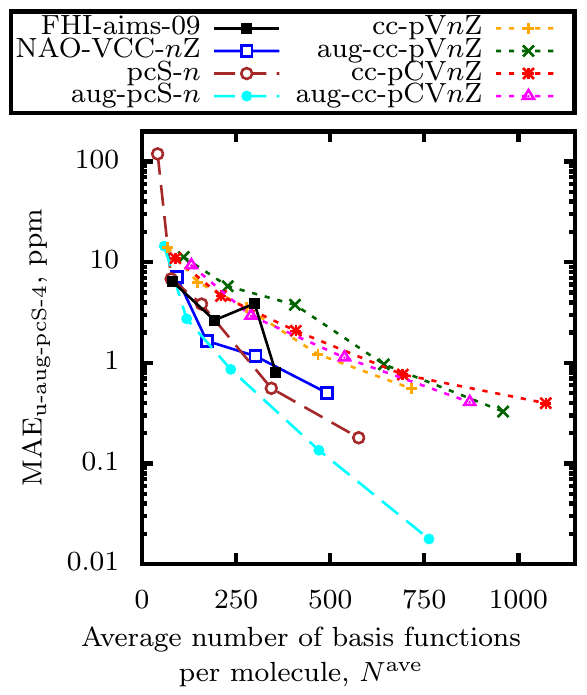}
  \caption{Mean average errors of chemical shifts (deviations from the values obtained with the u-aug-pcS-4 basis sets) for the test set of molecules shown in Fig.~\ref{fig:molecules}. Calculations were performed with the PBE functional. The levels shown for each basis set are the same as in Table~\ref{tab:basis_sets}.}
  \label{fig:abs_shift_err}
\end{figure}

In Figure~\ref{fig:abs_shift_err}, we next inspect the basis set convergence of chemical shifts for the same systems, a key quantity of interest in liquid state NMR methods.
The chemical shift scale [see Eq.~\eqref{eq:shift_def}] is defined by taking the shielding constant of a nucleus relative to a reference compound, which defines the ``zero-ppm'' shift for that nucleus. As reference compounds we use tetramethylsilane (TMS) (for $^{13}$C, $^1$H, and $^{29}$Si), CH$_3$NO$_2$ ($^{15}$N), SO$_4^{2-}$ ($^{33}$S), H$_3$PO$_4$ ($^{31}$P), H$_2$O ($^{17}$O) and CFCl$_3$ ($^{19}$F). The MAE of the chemical shifts are therefore affected by the respective errors of calculated shieldings for the reference compounds and for the test set. Chemical shieldings also depend on the solvent, but for the present purpose, solvent effects were neglected and the chemical shifts are calculated for rigid molecules in isolation.

As for the shieldings, we find that the NAO-VCC-$n$Z basis sets again exhibit good convergence behavior. Both the MAE of chemical shieldings and shifts have the same magnitude for a given basis set quality, indicating that these basis sets yield good stability and reliability. To obtain 1\,ppm accuracy for the chemical shifts, the largest NAO basis sets are needed (NAO-VCC-$n$Z with $n=4$ and $n=5$, or FHI-aims-09 tier 4), which is a considerable expense. The more affordable FHI-aims-09 tier 2 and NAO-VCC-3Z basis sets yield approximately 3\,ppm and 2\,ppm, respectively. As in Fig.~\ref{fig:abs_shield_err}, the tier 3 level of the FHI-aims-09 basis sets does not follow the same systematic trend as the other basis sets in this series. In fact, the choice of reference compounds (particularly the H$_2$O molecule) affects the behavior of tier 3. Finally, the GTO correlation consistent basis sets converge most slowly. Once again, for results very close to the basis set limit, the MAE could be biased towards the (aug)-pcS basis set series, since the selected reference is the largest uncontracted version of the same basis sets.

\subsection{Influence of system size on basis set convergence for shieldings}

\begin{figure}
  \centering
  \includegraphics[scale=1.15]{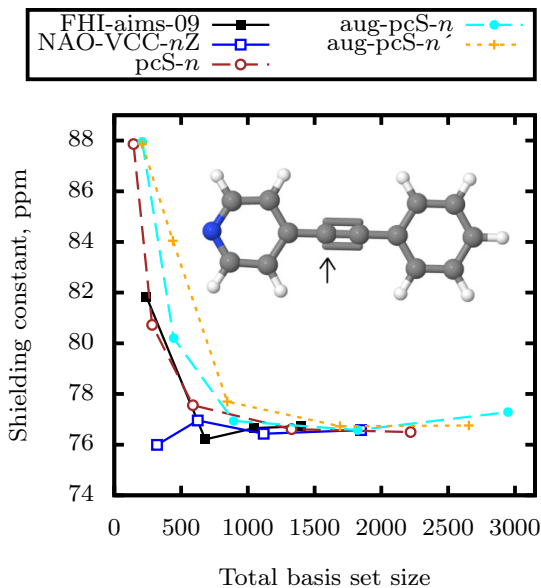}
  \caption{Convergence of the carbon shielding constant in PEP using the PBE functional. The levels shown for each basis set are the same as in Table~\ref{tab:basis_sets}. The aug-pcS-$n$' designation indicates the aug-pcS-$n$ basis sets, but after removal of any eigenvectors of the system overlap matrix with eigenvalues below 10$^{-5}$ from the basis set.}
  \label{fig:PPA_shield}
\end{figure}

Next, we investigate the influence of system size on basis set convergence for shieldings. We first compare the chemical shielding of the acetylene C atom (Fig.~\ref{fig:PPA_shield}) in the PEP molecule, where the basis set size for the PEP molecule is an order of magnitude larger than for molecules in the test set of Fig.~\ref{fig:molecules}.
In Fig.~\ref{fig:PPA_shield}, the correlation consistent basis sets were not considered since they were already shown to be less accurate compared to the other basis sets for a similar basis set size in Fig.~\ref{fig:abs_shield_err}. The FHI-aims-09 basis sets converge rapidly, reaching the same value obtained with the other basis sets at the tier~2 level. The NAO-VCC-$n$Z basis sets show very good accuracy (0.5\,ppm) for $n=2$ and remain essentially at the converged value for $n=3,4,5$.

One observation of note is that the aug-pcS-$n$ basis set shows a minor deviation from the common converged value for the shielding constant at the highest $n$. Specifically, for the aug-pcS-4 basis set (Fig.~\ref{fig:PPA_shield}, filled square on the far right) the obtained value slightly overshoots the consensus value of the other basis sets. As we show next, this overshooting is likely a consequence of basis set ill-conditioning \cite{golub13}, that is, a mutual interdependence of basis functions, where upon increase in system and basis set size the basis functions may become almost linearly dependent on each other. This near-linear dependence is reflected by very small eigenvalues of the system overlap matrix $S$ and will eventually lead to numerical inaccuracies due to the truncation errors imposed by the fixed numerical precision (here, double precision) of current computers.

In Table~\ref{tab:S_eigenvalues_PPA}, we inspect the eigenvalues of the PEP overlap matrix for four different basis set types.
\begin{table}
  \centering
  \caption{Lowest eigenvalues in the overlap matrix for different basis sets. $N_{\text{basis}}$ is the number of basis functions for the basis set in the PEP molecule and $L$ is the level of the basis set. $\min\lambda_S$ is the respective minimum value in the overlap matrix.}
  \begin{tabular}{rrlrrl}
    \toprule
    $L$ & $N_{\text{basis}}$ & $\min\lambda_S$ & $L$ & $N_{\text{basis}}$ & $\min\lambda_S$ \\
    \midrule
    \multicolumn{3}{c}{FHI-aims-09} & \multicolumn{3}{c}{NAO-VCC-$n$Z} \\
    \midrule
    1 & 241  & $1.1\times10^{-3}$ & 2Z & 320 & $1.7\times10^{-4}$ \\
    2 & 681  & $7.0\times10^{-5}$ & 3Z & 625 & $1.5\times10^{-5}$ \\
    3 & 1049 & $7.5\times10^{-6}$ & 4Z & 1119 & $2.6\times10^{-6}$ \\
    4 & 1399 & $7.5\times10^{-7}$ & 5Z & 1848 & $4.4\times10^{-7}$ \\
    \midrule
    \multicolumn{3}{c}{pcS-$n$} & \multicolumn{3}{c}{aug-pcS-$n$} \\
    \midrule
    0 & 144  &  $6.0\times10^{-3}$ & 0 & 209  &  $6.8\times10^{-6}$ \\
    1 & 283  &  $4.3\times10^{-4}$ & 1 & 445  &  $8.0\times10^{-7}$ \\
    2 & 588  &  $7.7\times10^{-5}$ & 2 & 893  &  $1.2\times10^{-9}$ \\
    3 & 1328 &  $1.8\times10^{-6}$ & 3 & 1822 &  $5.6\times10^{-11}$ \\
    4 & 2219 &  $6.9\times10^{-9}$ & 4 & 2948 &  $1.9\times10^{-12}$ \\
    \bottomrule
  \end{tabular}
  \label{tab:S_eigenvalues_PPA}
\end{table}
We see that the smallest eigenvalues in the aug-pcS-$n$ basis sets are significantly smaller than those of the FHI-aims-09, NAO-VCC-$n$Z, or pcS-$n$ basis sets. This is largely due to the presence of extended, diffuse augmentation functions in the aug-pcS basis sets \cite{ochsenfeld11}. If no eigenfunctions and eigenvalues are discarded this leads to numerical inaccuracies that are enough to influence the value of the shielding constant, as exemplified by aug-pcS-4 in Fig.~\ref{fig:PPA_shield}. The problem can be partially mitigated by discarding eigenvectors of the overlap matrix that correspond to eigenvalues smaller than a certain threshold. For a threshold of $10^{-5}$, the result is shown as aug-pcS-$n$' in Fig.~\ref{fig:PPA_shield}, indicating somewhat better numerical stability. However, as system sizes grow larger, discarding overlap matrix eigenvectors from the basis in this fashion introduces inaccuracies of its own, thus ultimately limiting the numerical precision that can be reached in a calculation \cite{mardirossian15},\cite{mardirossian16}. We note that no such threshold was applied to any molecule in the test set of small molecules.

\begin{figure}
  \centering
  \includegraphics[scale=1.3]{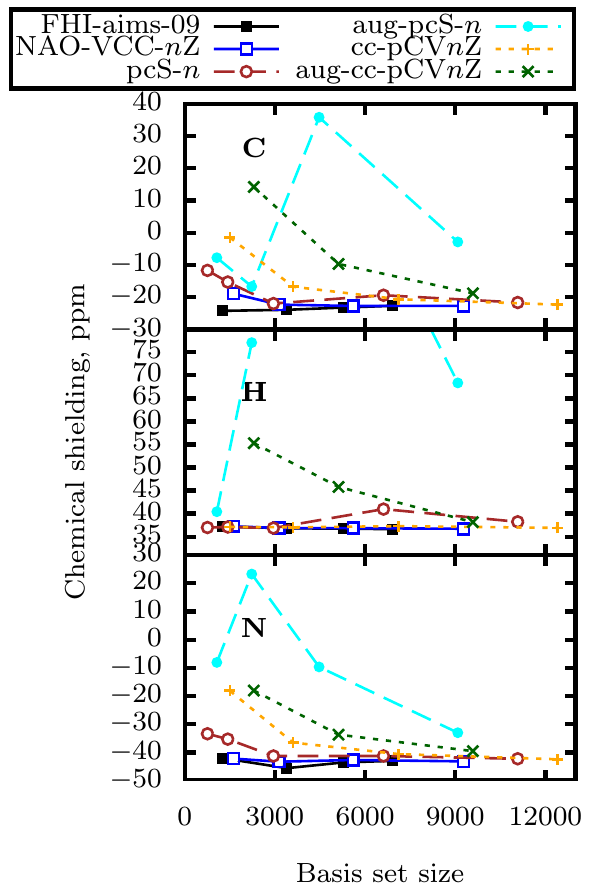}
  \caption{Convergence of the chemical shieldings of the PTC molecule (Fig.~\ref{fig:ir}). Calculations were performed using the PBE functional. The levels shown for each basis set are the same as in Table~\ref{tab:basis_sets} with the following exceptions: cc-pCV$n$Z: D--5, aug-cc-pCV$n$Z: D--Q, aug-pcS-$n$: 0--3. The augmented basis sets only employed eigenvectors of the system overlap matrix with eigenvalues greater than $10^{-5}$.}
  \label{fig:shield_ir}
\end{figure}

In Fig.~\ref{fig:shield_ir}, we next present results for a yet larger system, the PTC molecule.
In PTC, the ill-conditioning problem is significantly more severe than in the PEP molecule, as demonstrated by the overlap matrix elements in Table~\ref{tab:S_eigenvalues_Ir}.
\begin{table}
  \centering
  \caption{Lowest eigenvalues of the overlap matrix for the PTC molecule. $N_{\text{basis}}$ is the number of basis functions for the basis set in the PTC molecule and $L$ is the level of the basis set. $\min\lambda_S$ is the respective minimum value in the overlap matrix.}
  \begin{tabular}{rrlrrl}
    \toprule
    $L$ & $N_{\text{basis}}$ & $\min\lambda_S$ & $L$ & $N_{\text{basis}}$ & $\min\lambda_S$ \\
    \midrule
    \multicolumn{3}{c}{FHI-aims-09} & \multicolumn{3}{c}{NAO-VCC-$n$Z} \\
    \midrule
    1 & 1240 & $6.4\times10^{-4}$ & 2 & 1618 & $1.6\times10^{-4}$ \\
    2 & 3395 & $7.3\times10^{-5}$ & 3 & 3135 & $1.5\times10^{-5}$ \\
    3 & 5283 & $4.7\times10^{-6}$ & 4 & 5608 & $2.0\times10^{-6}$ \\
    4 & 6908 & $6.3\times10^{-7}$ & 5 & 9273 & $2.7\times10^{-7}$ \\
    \midrule
    \multicolumn{3}{c}{pcS-$n$} & \multicolumn{3}{c}{aug-pcS-$n$} \\
    \midrule
    0 & 756 & $4.6\times10^{-3}$   & 0 & 1069 & $1.0\times10^{-6}$ \\
    1 & 1435 & $2.4\times10^{-4}$  & 1 & 2232 & $3.5\times10^{-8}$ \\
    2 & 2952 & $2.2\times10^{-5}$  & 2 & 4469 & $3.5\times10^{-10}$ \\
    3 & 6612 & $3.7\times10^{-7}$  & 3 & 9085 & $2.1\times10^{-13}$ \\
    4 & 11074 & $1.2\times10^{-9}$ & 4 & 14739 & $1.4\times10^{-14}$ \\
    \midrule
    \multicolumn{3}{c}{cc-pCV$n$Z} & \multicolumn{3}{c}{aug-cc-pCV$n$Z} \\
    \midrule
    D & 1500  & $5.2\times10^{-5}$ & D & 2297 &$1.7\times10^{-7}$ \\
    T & 3602  & $9.3\times10^{-6}$ & T & 5119 &$1.6\times10^{-8}$ \\
    Q & 7115  & $1.3\times10^{-6}$ & Q & 9588 &$1.3\times10^{-9}$ \\
    5 & 12405 & $1.1\times10^{-7}$ & 5 & 16070 &$7.8\times10^{-11}$ \\
    6 & 19838 & $3.3\times10^{-8}$ \\
    \bottomrule
  \end{tabular}
  \label{tab:S_eigenvalues_Ir}
\end{table}
Inclusion of augmentation functions leads to extremely small eigenvalues in the overlap matrix. In some cases the difference between augmented and non-augmented versions is several orders of magnitude. As a result, the calculations become inaccurate upon basis set size increase and we were unable to compute the shieldings for some of the augmented GTO basis sets. Consequently, the shieldings shown in Fig.~\ref{fig:shield_ir} were obtained with GTO basis sets that were reduced by discarding any eigenfunctions of the overlap matrix that corresponded to eigenvalues smaller than $10^{-5}$ in the overlap matrix. That threshold was applied to the cc-pCV$n$Z, aug-cc-pCV$n$Z, pcS-$n$, and aug-pcS-$n$, but not the FHI-aims-09 and NAO-VCC-$n$Z basis sets. We see that even with the basis reduction technique, in the present implementation some of the augmented GTO basis sets did not yield results that are fully numerically stable due to severe ill-conditioning. Using the full basis sets (not shown) either had no effect or increased the numerical problems for the cc-pCV$n$Z, aug-cc-pCV$n$Z, pcS-$n$, and aug-pcS-$n$ basis sets.

Overall, the NAO (FHI-aims-09 tiers 1,2,4 and NAO-VCC-$n$Z) basis sets show similar convergence behavior as the polarization consistent basis sets for small molecules for the calculation of chemical shieldings and shifts. The correlation consistent basis sets lead to the slowest convergence in most cases. For larger systems, the NAO basis sets are numerically more stable than the GTO basis sets, especially with augmentation functions, in that they suffer less from basis set ill-conditioning.

\subsection{Magnetizabilities}

\begin{figure}
  \centering
  \includegraphics[scale=1.35]{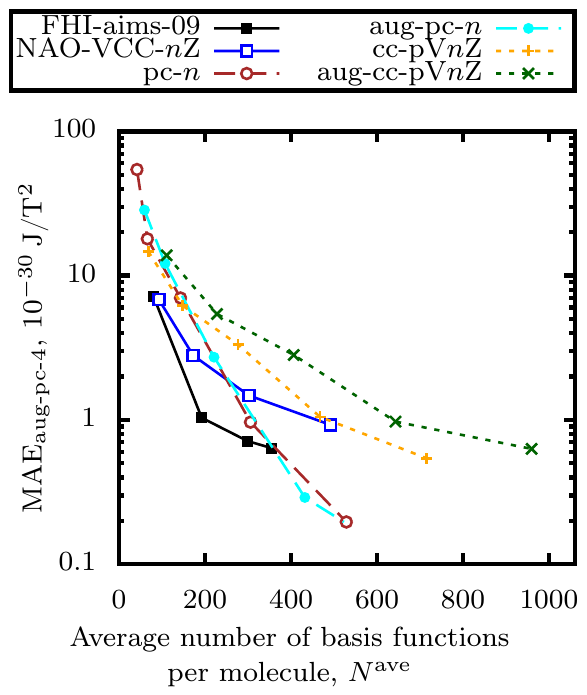}
  \caption{Mean average errors of calculated magnetizability (deviations from the values obtained with the aug-pc-4 basis sets) for the test set of molecules shown in Fig.~\ref{fig:molecules}. Calculations were performed with the PBE functional. The levels shown for each basis set are the same as in Table~\ref{tab:basis_sets}.}
  \label{fig:abs_magnet_err}
\end{figure}

The reference values for estimating the magnetizability errors were determined using the aug-pc-4 basis sets \cite{jensen01}. Uncontraction of the basis sets does not lead to any noticeable improvement. This is expected because the extra flexibility in the core region should not benefit the magnetization across the molecule as much as the chemical shieldings. It is seen from the MAE of magnetizability (Fig.~\ref{fig:abs_magnet_err}) that the FHI-aims-09 basis sets converge the fastest, followed by the NAO-VCC-$n$Z, aug-pc-$n$, and aug-cc-pV$n$Z basis sets.
The MAEs could again be biased towards the larger (aug)-pc-$n$ basis sets due to the chosen reference.

\begin{figure}
  \centering
  \includegraphics[scale=1.25]{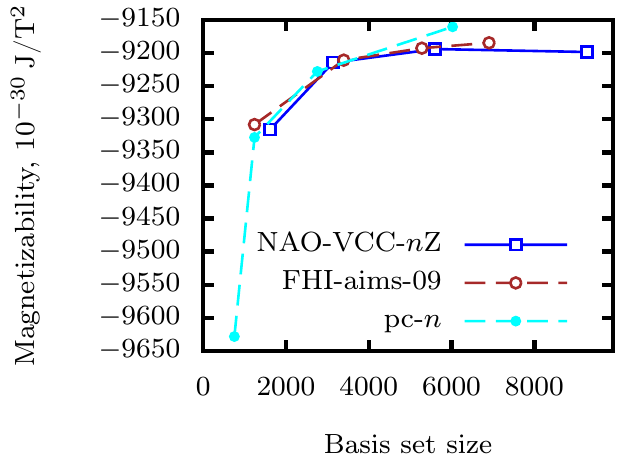}
  \caption{Convergence of the magnetizability of the PTC molecule using the PBE functional. The levels shown for each basis set are the same as in Table~\ref{tab:basis_sets} with the following exception, pc-$n$: 0--3.}
  \label{fig:magnet_ir}
\end{figure}

Regarding the medium-scale PTC molecule, we encountered similar issues as for the chemical shieldings. Due to ill-conditioning problems, we were unable to obtain converged results with the aug-cc-pV$n$Z or the aug-pc-$n$ basis sets. Even the pc-$n$ basis sets do not appear to converge to the exact same magnetizability as the NAO basis sets, as is demonstrated in Fig.~\ref{fig:magnet_ir}. 
The magnetizabilities were obtained with the full pc-$n$ basis sets. Reducing the basis set size as was done for shieldings did not help here and the numerical errors actually increased (not shown). The NAO basis sets do not suffer as much from ill-conditioning as the augmented GTO basis sets and seem to approach a consistent basis set limit.

\subsection{\textit{J}-couplings}

\begin{figure}
  \centering
  \includegraphics{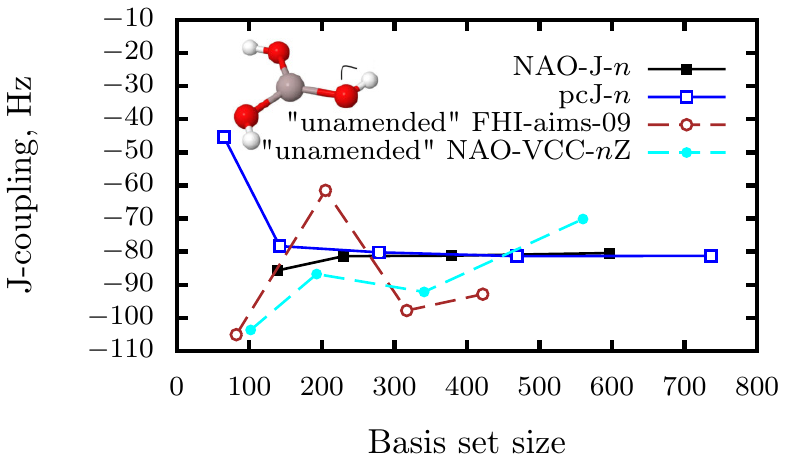}
  \caption{Basis set convergence of the $^{17}$O-$^1$H \textit{J}-coupling of Al(OH)$_3$ using the NAO and GTO basis sets. ``Unamended'' FHI-aims-09 or NAO-VCC-$n$Z means that the additional near-nuclear Gaussian s-type primitive functions from the pc-J basis sets were not used.}
  \label{fig:aloh3_oh}
\end{figure}

We next examine the convergence of \textit{J}-couplings with basis set size and choice.
In previous studies of some of the GTO (aug-cc-pV$n$Z, cc-pCV$n$Z, pc-$n$) basis set types, the convergence of \textit{J}-couplings with basis set size was shown to be somewhat erratic \cite{deng06},\cite{jensen06}. These basis sets are therefore not considered here. For the example of the $^{17}$O-$^1$H \textit{J}-coupling of Al(OH)$_3$, Fig.~\ref{fig:aloh3_oh} shows that the plain FHI-aims-09 and NAO-VCC-$n$Z basis sets (called ``unamended'' in the figure) are affected by the same erratic behavior.
The reason is primarily due to the Fermi contact term, whose matrix are evaluated using a single point at the nucleus [Eq.~\eqref{eq:FC}], which requires extreme flexibility of the basis sets in the core region. In other words, for a highly localized perturbation near the nucleus, the standard basis sets do not have enough variational freedom to accurately describe the first-order response. The portion of the Hilbert space spanned by a standard general-purpose basis set may be sufficient to accurately yield the zero-order orbitals for computing, e.g., the ground state energy, but insufficient to yield accurate first-order orbitals.

Often a solution is to increase the flexibility of the basis sets by uncontracting and increasing the number of orbitals (by adding, most importantly, tight s functions). Following this recipe, Jensen developed the pcJ-$n$ basis sets \cite{jensen06}, which we take as reference for \textit{J}-couplings in this work. Fig.~\ref{fig:aloh3_oh} demonstrates that the $^{17}$O-$^1$H \textit{J}-coupling, calculated using the pcJ-$n$ basis sets, exhibits smooth numerical convergence to the basis set limit. For the NAOs there is no immediate analogue to uncontraction. To achieve added flexibility, we therefore take the innermost s-type GTOs from the \textit{uncontracted} pcJ-$n$ basis sets and use them to augment the NAO-VCC-$n$Z basis sets, resulting in what we call the ``NAO-J-$n$'' basis sets. The number of these additional functions was empirically determined for each element. Four tight s-type GTOs for H, B, F, Al, Si, P, S, Cl, six for C and N, and seven for O are sufficient to achieve the numerical convergence properties outlined below.

\begin{figure}
  \centering
  \includegraphics{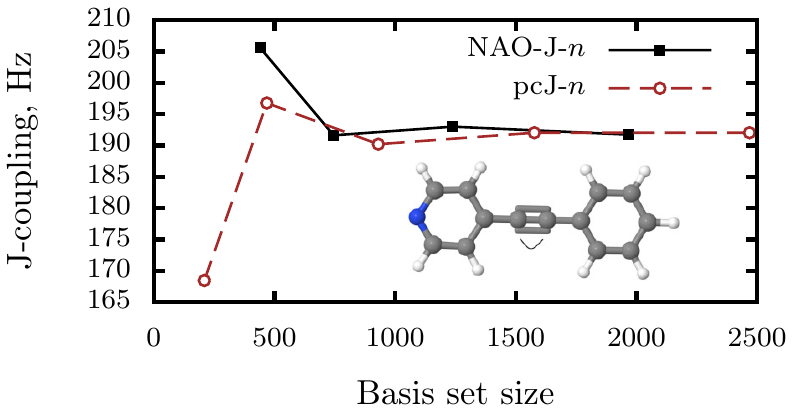}
  \caption{Basis set convergence of the $^{13}$-C$^{13}$C \textit{J}-coupling of PEP using the NAO-J-$n$ and the GTO pcJ-$n$ basis sets.}
  \label{fig:ppa_cc}
\end{figure}

In Fig.~\ref{fig:aloh3_oh}, smooth convergence to the basis set limit is recovered for the $^{17}$O-$^1$H \textit{J}-coupling when using the NAO-J-$n$ basis sets. A second example, the $^{13}$C-$^{13}$C \textit{J}-coupling in the larger PEP molecule as shown in Fig.~\ref{fig:ppa_cc}, corroborates the systematic and similar convergence behavior of the pcJ-$n$ and NAO-J-$n$ basis sets for $J$-couplings.
Although we did not pursue a broader comparison to experimental reference data in this study, it is reassuring that the converged DFT-PBE value of 192\,Hz agrees reasonably well with the experimental value of 185\,Hz \cite{zhou17}. We also attempted to add tight s-type primitive GTOs to the FHI-aims-09 basis sets to see if a similar convergence improvement would result. However, the NAO-J-$n$ basis sets remained superior and we therefore restrict the remaining discussion to this approach.

\begin{figure}
  \centering
  \includegraphics[scale=1.35]{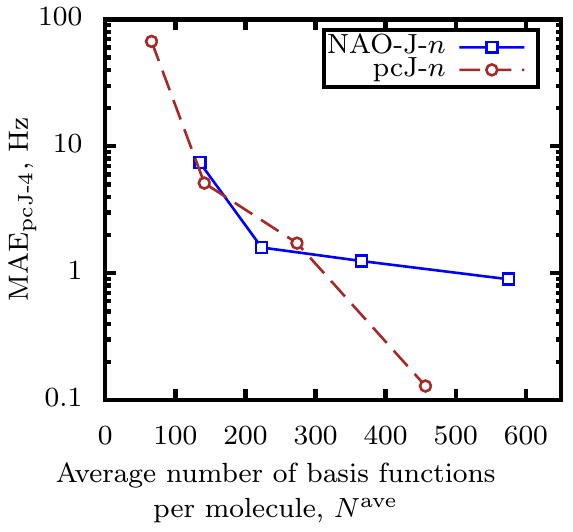}
  \caption{Mean average error of calculated \textit{J}-couplings, averaged over the 46 \textit{J}-couplings listed in Table~\ref{tab:j_couplings}, using the values obtained with the pcJ-4 basis set as a reference. Detailed convergence studies are shown in the Supporting Information (Figs.~S74--S91).}
  \label{fig:abs_j}
\end{figure}

Table~\ref{tab:j_couplings} lists all pairs of atoms for which the $J$-couplings were calculated.
\begin{table}
  \centering
  \caption{Test set of molecules and atom pairs for which \textit{J}-couplings were calculated and averaged over in Fig.~\ref{fig:abs_j}.}
  \begin{tabular*}{\columnwidth}{rl}
    \toprule
    Molecule & Selected atom pair(s) \\
    \midrule
    Al$_2$Cl$_6$: & Al-Cl, Al-Al \\
    Al(OH)$_3$: & Al-O, Al-H, O-H \\
    B$_2$H$_6$: & B-B, B-H \\
    BCl$_3$: & B-Cl, Cl-Cl \\
    C$_6$H$_6$: & C-C, C-H, H-H \\
    CF$_4$: & C-F, F-F \\
    CFCl$_3$: & C-F, C-Cl, Cl-Cl \\
    CH$_3$NO$_2$: & C-N, N-O, O-O \\
    H$_2$C$_2$O: & C-C, C-O $^2$J(C-O) \\
    H$_2$O: & O-H, H-H \\
    H$_2$SO$_4$: & S-O1, S-O2, S-H \\
    H$_3$PO$_4$: & P-O1, P-O2, P-H \\
    HFCO: & C-F, C-O, F-O \\
    NH$_3$: & N-H, H-H \\
    PClF$_2$: & P-F, P-Cl, F-F \\
    SiF$_4$: & Si-F, F-F \\
    SO$_4^{2-}$: & S-O, O-O \\
    TMS: & Si-C \\
    \bottomrule
  \end{tabular*}
  \label{tab:j_couplings}
\end{table}
Figure~\ref{fig:abs_j} shows the mean average deviations of 46 \textit{J}-couplings (listed in Table~\ref{tab:j_couplings}) as a function of basis set size and type for the NAO-J-$n$ and the GTO pcJ-$n$ basis sets, using the large pcJ-4 basis sets as a reference.
Just like for shieldings and magnetizabilities, it is possible that the use of pcJ-4 as a reference introduces a slight bias in favor of the other pcJ basis sets (particularly pcJ-3). In any case, numerical convergence to less than 1~Hz on average is demonstrated for both the NAO-J-$n$ and the GTO pcJ-$n$ basis sets.

We also attempted to conduct the same convergence studies for the larger aug-pcJ-$n$ basis sets. However, we found that basis set ill-conditioning already led to numerical problems for the test set of small molecules. In particular, the DFPT cycle became unstable for some systems. In these tests, the aug-pcJ-$n$ basis sets were not reduced by discarding eigenfunctions of the overlap matrix that corresponded to small eigenvalues in the overlap matrix. While it is possible that this reduction would have remedied the numerical convergence problems, we did not pursue the aug-pcJ-$n$ further in this study.

\subsection{Larger scale calculations}

We finally describe and validate how shieldings and \textit{J}-couplings can be computed efficiently for very large systems. For some scenarios, it is sufficient to obtain shieldings and \textit{J}-couplings only for a limited number of atoms and atom pairs in an otherwise much larger system. For example, this is the case when aiming to describe the essential nuclear spin physics near the center of a polarization transfer catalyst as shown in Fig.~\ref{fig:ir}.

As a simple example, we calculate a specific atom's carbon shielding constant and a specific atom pair's CH \textit{J}-coupling constant in the DNA segment shown in Fig.~\ref{fig:dna}.
In Table~\ref{tab:dna_shield}, we show results for the C shielding. Here, ``light'' refers to FHI-aims-09 with the minimal basis and the s, p, and d basis functions of tier 1, and relatively sparse numerical grids.
\begin{table}
  \centering
  \caption{Convergence study of the selected C atom shielding constant in the DNA segment (see Fig.~\ref{fig:dna}). Calculations were performed with the PBE density functional. The NAO-VCC-2Z, NAO-VCC-3Z, NAO-VCC-4Z, FHI-aims-09 based ``light'', or ``tight'' basis sets were used on all atoms in the calculations.}
  \begin{tabular*}{\columnwidth}{rrr}
    \toprule
    Basis set & Basis size & $\sigma_C$ \\
    \midrule
    NAO-VCC-2Z & 14\,972 & 96.0\,ppm \\
    NAO-VCC-3Z & 28\,930 & 97.6\,ppm \\
    NAO-VCC-4Z & 51\,672 & 97.2\,ppm \\
    Light & 11\,230 & 99.4\,ppm \\
    Tight & 31\,380 & 96.9\,ppm \\
    \bottomrule
  \end{tabular*}
  \label{tab:dna_shield}
\end{table}
``Tight'' refers to FHI-aims-09 with tier 1 and, depending on the element, all or some of the tier 2 basis functions, and tighter settings for the numerical grids. (Full specifics of the light/tight basis functions for light elements are given in Ref.~\cite{elephantSI}). According to Table~\ref{tab:dna_shield}, the shielding constant is essentially converged at the NAO-VCC-3Z basis set or FHI-aims-09:tight. This degree of basis set convergence is somewhat better than indicated by the MAE for small molecules in Fig.~\ref{fig:abs_shield_err}. At this point, errors introduced by the overall approximations made at this level of theory (vibrations, solvent effects, etc.) are expected to be larger or comparable to the basis set error.

For the \textit{J}-coupling calculation, we demonstrate a ``hybrid'' basis set approach in Table~\ref{tab:dna_j}, using a relatively simple general-purpose basis set for most atoms while reserving a high-accuracy, computationally more expensive basis set only for the specific atoms for which shieldings and/or \textit{J}-couplings are required.
\begin{table}
  \caption{Convergence study of the CH \textit{J}-coupling constant in the DNA segment (see Fig.~\ref{fig:dna} for positions). In all calculations, the NAO-J-5 basis set was placed at the C and H atoms. The other atoms used basis sets from the NAO-VCC-$n$Z series or FHI-aims-09:light/tight.}
  \centering
  \begin{tabular*}{\columnwidth}{rrr}
    \toprule
    Basis set & Basis size & J$_{\text{CH}}$ \\
    \midrule
    NAO-VCC-2Z/NAO-J-5 & 15\,108 & 145.66\,Hz \\
    NAO-VCC-3Z/NAO-J-5 & 29\,041 & 145.07\,Hz \\
    NAO-VCC-4Z/NAO-J-5 & 51\,742 & 144.98\,Hz \\
    Light/NAO-J-5 & 11\,373 & 145.46\,Hz \\
    Tight/NAO-J-5 & 31\,488 & 145.24\,Hz \\
    \bottomrule
  \end{tabular*}
  \label{tab:dna_j}
\end{table}
Specifically, we placed the largest J-optimized basis sets, the NAO-J-5, on the two C/H atoms of interest, and the NAO-VCC-$n$Z or FHI-aims-09 basis sets on other atoms. Interestingly, Table~\ref{tab:dna_j} shows that the \textit{J}-coupling constant converges already with the NAO-VCC-2Z/NAO-J-5 or FHI-aims-09:light/NAO-J-5 basis set, i.e., using the smallest members of the NAO-VCC-$n$Z or FHI-aims-09 series.
Comparing the basis set sizes between Tables~\ref{tab:dna_shield} and \ref{tab:dna_j}, we note that the sizes have increased only slightly due the inclusion of the NAO-J-5 basis sets. Thus, the calculations are relatively cheap if only a few \textit{J}-couplings are required.

\section{Conclusions}

In this paper, we discuss the computational formalism for and basis set convergence of calculated NMR chemical shieldings, magnetizabilities, and \textit{J}-couplings, using numeric atom-centered orbitals (NAOs) as basis functions. The NAO basis sets are designed to be scalable to large systems at high numerical accuracy for total energy calculations, motivating their use for magnetic response properties aimed at large molecules. We assess two types of general-purpose NAO basis sets: FHI-aims-09 \cite{blum09} and NAO-VCC-$n$Z \cite{zhang13}. For magnetizabilities, the NAO basis sets exhibit faster convergence to the basis set limit than several commonly used GTO basis sets. For shieldings, the NAO-VCC-$n$Z perform better than the correlation consistent GTO basis sets and they are on par with the polarization consistent basis sets optimized for shieldings, pcS-$n$ \cite{jensen08}. The calculation of \textit{J}-couplings is known to be challenging, primarily due to the Fermi contact term, which requires the basis sets to be flexible near the atomic nucleus. We follow the successful recipe of Jensen's pcJ-$n$ \cite{jensen06} basis sets, by adding the same tight s orbitals used in the pcJ-$n$ basis sets to the NAO-VCC-$n$Z basis sets. The resulting new basis sets, called NAO-J-$n$, are shown to be accurate and have similar convergence properties as the pcJ-$n$ basis sets.

We summarize the suitability of the NAO basis sets considered in this work for magnetic response calculations as follows:
\begin{itemize}
\item For shieldings, the mean average errors are lower for the NAO-VCC-$n$Z than the FHI-aims-09 basis sets. For high-accuracy shielding calculations, we therefore recommend the NAO-VCC-4Z or 5Z basis sets. Furthermore, a test calculation for a large scale molecule (the 1,052 atom DNA segment in Table \ref{tab:dna_shield}) indicates that in systems closer to a condensed-phase environment, smaller basis sets may additionally benefit from the overall increased density of basis functions in the system.
\item Magnetizabilities convergence quickly using both the NAO-VCC-$n$Z and FHI-aims-09 basis sets, the latter exhibiting slightly faster convergence.
\item For \textit{J}-couplings, the NAO-J-$n$ basis sets show good convergence and accuracy that is comparable with the pcJ-$n$ basis sets \cite{jensen06} down to approximately 1~Hz. As demonstrated in Table~\ref{tab:dna_j}, it is sufficient to place a dense NAO-J-$n$ basis set on the atoms of interest and cheaper basis sets on the rest. We recommend the NAO-J-4 or NAO-J-5 basis sets for accurate calculations of \textit{J}-couplings.
\item Unlike for shieldings or for magnetizabilities, the \textit{J}-coupling calculations should use a dense radial integration grid near the nucleus of interest. In our present implementation in FHI-aims, this is only possible by increasing the number of radial grid shells uniformly for the entire atom of interest (approximately four times the number of grid shells required for a normal total-energy calculation using FHI-aims' ``tight'' species defaults). Future work on the grids may help reduce this requirement by restricting the denser integration grids only to the region near the nucleus.
\item Finally, in practical calculations, the NAO basis sets assessed here seem to suffer somewhat less from possible basis set ill-conditioning problems than the ``augmented'' GTO basis set types (when assessed within the exact same numerical framework). Due to the rather stringent basis set size requirements for shieldings and $J$-couplings, this is an issue that will tend to arise in larger systems such as the PTC molecule tested in this study.
\end{itemize}
Overall, this paper lays the ground work to place the otherwise very promising NAO basis sets in the context of the established GTO basis sets in the literature for NMR. This shows that, for light elements, the NAO basis sets can be used essentially on par with the best available reference basis sets of which we are aware, Jensen's pcS-$n$ and pcJ-$n$ basis sets. Beyond the present work, we expect that our new implementation and benchmarks (which are here available and documented) will also serve as the foundation for important further developments, such as improved numerical representations of the atomic part of the magnetic response near the nucleus.

\section*{Acknowledgement }

Acknowledgment is made to the Donors of the American Chemical Society Petroleum Research Fund for partial support of this research. Part of this work was also supported by the National Science Foundation under Grant Number 1450280.

\appendix
\section*{Appendix}
\subsection*{GGA exchange-correlation kernel}

Given a GGA exchange-correlation energy as a functional of the spin densities and the spin density gradients,
\begin{equation}
  E_{\text{xc}} \equiv E_{\text{xc}}[n_\alpha, n_\beta, \bm\nabla n_\alpha, \bm\nabla n_\beta],
\end{equation}
the exchange-correlation potential, for the spin channel $\alpha$, can be written as
\begin{equation}
  \begin{split}
    V_{\text{xc},\alpha}(\bm r) =& \frac{\delta E_{\text{xc}}}{\delta n_\alpha} + \frac{\delta E_{\text{xc}}}{\delta \bm\nabla n_\alpha} \bm\nabla \\
    =& \frac{\partial E_{\text{xc}}}{\partial n_\alpha} + 2\frac{\partial E_{\text{xc}}}{\partial \gamma_{\alpha\alpha}}\bm\nabla n_\alpha \bm\nabla + \frac{\partial E_{\text{xc}}}{\partial \gamma_{\alpha\beta}}\bm\nabla n_\beta \bm\nabla, \\
    \gamma_{\sigma\sigma'} =& \bm\nabla n_\sigma\bm\nabla n_{\sigma'}.
  \end{split}
\end{equation}
Before finding the second derivative of $E_{\text{xc}}$, it is convenient to first express the unperturbed matrix elements of the xc-potential in terms of the scalar and vectorial parts, $u_\alpha^{(0)}$ and $\bm v_\alpha^{(0)}$:
\begin{equation}
  \begin{split}
    & \Braket{\phi_i | V_{\text{xc},\alpha} | \phi_j} = \int u_\alpha^{(0)} \Omega_{ij}(\bm r) \,\mathrm{d}\bm r \\
    &+ \int \bm v_\alpha^{(0)} \bm\nabla\Omega_{ij}(\bm r) \,\mathrm{d}\bm r, \\
& u_\alpha^{(0)} = \frac{\partial E_{\text{xc}}}{\partial n_\alpha}, \\
& \bm v_\alpha^{(0)} = 2\frac{\partial E_{\text{xc}}}{\partial \gamma_{\alpha\alpha}}\bm\nabla n_\alpha + \frac{\partial E_{\text{xc}}}{\partial \gamma_{\alpha\beta}}\bm\nabla n_\beta, \\
& \Omega_{ij}(\bm r) = \phi_i(\bm r)\phi_j(\bm r).
  \end{split}
\end{equation}
The matrix elements of the first-order xc-potential are then given by
\begin{equation}
\label{eq:uv_def}
  \begin{split}
    V_{\text{xc};ij}^{(1)} = & \Bigg< \phi_i \Bigg| \frac{\partial V_{xc,\alpha}}{\partial n_\alpha} n_\alpha^{(1)} + \frac{\partial V_{xc,\alpha}}{\partial n_\beta} n_\beta^{(1)} \\
    & + \sum_{\sigma\sigma'}\frac{\partial V_{xc,\alpha}}{\partial\gamma_{\sigma\sigma'}} \bm\nabla n_\sigma\bm\nabla n_{\sigma'}^{(1)} \Bigg| \phi_j \Bigg> \\
    &= \int u_\alpha^{(1)} \Omega_{ij}(\bm r) \,\mathrm{d}\bm r + \int \bm v_\alpha^{(1)} \bm\nabla\Omega_{ij}(\bm r) \,\mathrm{d}\bm r,
  \end{split}
\end{equation}
where, for spin channel $\alpha$,
\begin{equation}
  \begin{split}
    u_\alpha^{(1)} =& E_{\text{xc}}^{(\alpha\alpha)}n_\alpha^{(1)}+ E_{\text{xc}}^{(\alpha\beta)}n_\beta^{(1)} + \sum_{\sigma\sigma'} E_{\text{xc}}^{(\alpha,\sigma\sigma')} \gamma_{\sigma\sigma'}^{(1)}, \\
    \bm v_\alpha^{(1)} =& 2\sum_\sigma E_{\text{xc}}^{(\sigma,\alpha\alpha)}\bm\nabla n_\alpha n_\sigma^{(1)} + \sum_\sigma E_{\text{xc}}^{(\sigma,\alpha\beta)}\bm\nabla n_\beta n_\sigma^{(1)} \\
    &+ 2\sum_{\sigma\sigma'} E_{\text{xc}}^{(\alpha\alpha,\sigma\sigma')}\bm\nabla n_\alpha\gamma_{\sigma\sigma'}^{(1)} \\
    &+ \sum_{\sigma\sigma'} E_{\text{xc}}^{(\alpha\beta,\sigma\sigma')}\bm\nabla n_\beta\gamma_{\sigma\sigma'}^{(1)} \\
    &+ 2E_{\text{xc}}^{(\alpha\alpha)}\bm\nabla n_\alpha^{(1)} + E_{\text{xc}}^{(\alpha\beta)}\bm\nabla n_\beta^{(1)}.
  \end{split}
\end{equation}
The short-hand notation for the energy derivatives is
\begin{equation}
  \begin{split}
    & E_{\text{xc}}^{(\sigma_1\sigma_2)} = \frac{\partial E_{\text{xc}}}{\partial\gamma_{\sigma_1\sigma_2}}; \quad E_{\text{xc}}^{(\sigma_1,\sigma_2\sigma_3)} = \frac{\partial^2 E_{\text{xc}}}{\partial n_{\sigma_1} \partial \gamma_{\sigma_2\sigma_3}} \\
    & E_{\text{xc}}^{(\sigma_1\sigma_2,\sigma_3\sigma_4)} = \frac{\partial^2 E_{\text{xc}}}{\partial\gamma_{\sigma_1\sigma_2}\partial\gamma_{\sigma_3\sigma_4}}.
\end{split}
\end{equation}
First-order density and density gradient are evaluated according to
\begin{equation}
  \begin{split}
  n^{(1)}(\bm r) =& \sum_{ij} \phi_i(\bm r) n_{ij}^{(1)} \phi_j(\bm r), \\
  \bm\nabla n^{(1)}(\bm r) & \\
  =& \sum_{ij} [\bm\nabla\phi_i(\bm r) n_{ij}^{(1)} \phi_j(\bm r) + \phi_i(\bm r) n_{ij}^{(1)} \bm\nabla\phi_j(\bm r)].
  \end{split}
\end{equation}
If the perturbation behind the first-order density is a magnetic moment, then $V_{\text{xc};ij}^{(1)} = V_{\text{xc};ij}^{\bm\mu_A}$ as in Eq.~\eqref{eq:U_j}.

The expressions shown here are applicable in a spin-polarized calculation using an LDA or a GGA functional. In LDA, all the gradient containing terms vanish. In case of \textit{J}-couplings, if the unperturbed electronic structure is not spin-polarized, the expressions simplify somewhat and the cost of calculation of the matrix elements $V_{\text{xc};ij}^{(1)}$ reduces by up to a factor of one half. The reason is that, even though we start with a non-spin-polarized system, the first-order density is always spin-polarized. This is in contrast to, e.g., phonons by DFPT, where a non-spin-polarized system remains non-spin-polarized after switching on the perturbation and the calculation of the matrix elements is significantly cheaper than in the spin-polarized case.


\end{document}